\documentclass[aps,pra,onecolumn,longbibliography,superscriptaddress,final,floatfix, 11pt,tightenlines]{revtex4-2}
\usepackage{amsfonts}
\usepackage{amsmath}
\usepackage{amssymb}
\usepackage{latexsym}
\usepackage[export]{adjustbox}
\usepackage{array}
\newcolumntype{P}[1]{>{\centering\arraybackslash}p{#1}}
\newcolumntype{M}[1]{>{\centering\arraybackslash}m{#1}}
\usepackage{caption}
\usepackage[colorlinks,citecolor=red,urlcolor=blue,bookmarks=false,hypertexnames=true]{hyperref} 
\usepackage[usenames, dvipsnames]{color}
\usepackage[svgnames]{xcolor}
\usepackage{color}
\usepackage{bm}
\usepackage{relsize}
\usepackage{float}
\usepackage{subfloat}
\usepackage{dcolumn}
\usepackage{epsfig}
\usepackage{subcaption}
\usepackage{graphicx}
\usepackage{multirow}
\usepackage{makecell}
\usepackage{mathtools}
\usepackage[bbgreekl]{mathbbol}

\usepackage{footnote}
\usepackage{balance}

\begin{document}

\title{Decoherence and the ultraviolet cutoff:\\ non-Markovian dynamics of a charged particle in a magnetic field}

\author{Suraka Bhattacharjee}
 \affiliation{Raman Research Institute, Bangalore-560080, India} 
\author{Koushik Mandal}
\affiliation{Institute of Physics, Sachivalaya Marg, Bhubaneswar 751005, India
}    
 \author{Supurna Sinha}
 \affiliation{Raman Research Institute, Bangalore-560080, India}

\date{\today}
\begin{abstract}
We derive a non-Markovian master equation for a charged particle in a magnetic field coupled to a bath and 
study decoherence by analysing the temporal decay of the off-diagonal elements
of the reduced density matrix in the position basis. 
The coherent oscillations characterised by the cyclotron frequency get suppressed as a result of decoherence due to coupling with the environment. 
We consider an Ohmic bath with 
three distinct models for the high-frequency cutoff for the spectral density of the bath and compare the three cases. As expected, the three cutoff models converge in the limit of the uppermost frequency of the bath tending to infinity. We notice
a dramatic slowing down of loss of coherence in the low-temperature limit dominated by zero point quantum fluctuations compared to the high-temperature classical limit dominated by 
thermal fluctuations. We also go beyond the Ohmic model and study 
super-Ohmic and sub-Ohmic baths with the spectral densities deviating from a linear dependence on the frequency.
Our results are testable in a state of the art cold atom laboratory. 
\end{abstract}

\maketitle
\section{Introduction}
Decoherence is central to understanding 
the transition from a quantum to a classical world, which has an extensive application in the modern quantum era, encompassing the development of quantum qubits and superconducting quantum processors \cite{Aaronson,Akama,Fischer2}. There has been a lot 
of work in this area \cite{Zeh,Schlosshauer,Zurek1,Zurek2,Caldeira1,Caldeira2} and the exact master equation for Brownian motion in the presence of non-local dissipation and colored noise was first derived in \cite{Unruh, HuPaz}. It has been studied in a variety of 
contexts. Some of the studies have focused on the fundamental aspects of the
quantum to classical transition \cite{Brun,Xiong,Zurek3}.
Many researchers have applied the concept of decoherence to different systems in the realms of 
quantum optics \cite{Orszag} and condensed matter
physics, including superconducting circuits \cite{Cucchietti,Devoret,Makhlin}. The effect of zero point fluctuations has been explored extensively in the context of quantum Brownian motion \cite{Sinhadecoherence,Ford,Mohanty,Lombardo}.
The effect of zero point fluctuations
contained in the Fluctuation-Dissipation theorem (FDT) and the Aharonov-Bohm effect limits the phase coherence time in mesoscopic systems, as was studied theoretically in a quasi-1D gold wire and later in mesoscopic rings designed on InGaAs/InAlAs heterostructure \cite{Mohanty2,Ren,Giulio}. The outcomes show the emergence of the power law decay of coherence at long times, which is also consistent with the experimental results \cite{Hanson}. There have been multiple researches to increase the coherence time using the valence holes in quantum dots of GaAs and other carbon-based materials like graphene, nanotubes \cite{Fischer} and the dynamics of single-spin diamond has also been reported \cite{Hanson2}. These studies are based on the occurrence of decoherence at absolute zero temperature, arising due to the non-commutativity of the unperturbed system Hamiltonian and the interaction Hamiltonian. The decay of coherence is also affected by the nature of the spectral density functions. The non-exponential decay of unstable quantum states has been investigated in some other recent studies using the exact dynamics of quantum Brownian motion \cite{Beau}. The dynamic and instantaneous decoherence, signifying the decoherence between the components of the initial state and components of the current states, were studied in a semi-classical approach \cite{Polonyi}. The results show a double exponential trend in the presence of a harmonic oscillator potential \cite{Polonyi}.  The Born-Markov master equation has also been studied for a quantum Brownian particle coupled to a bath of two-level systems \cite{Schlosshauertwolevel}. While most of these studies have focused on the Markovian version of the master equation, quite a few researches have been based on the non-Markovian generalized form of the Hu-Paz-Zhang (HPZ) master equation and explored the effects of non-Markovianity on the loss of coherence in the presence of system-bath interaction \cite{HuPaz}. The non-Markovianity is primarily manifested in the coefficients of the terms in the master equation, which are complicated and rather difficult to compute for a generalized spectral bath density. In this context, some authors have derived the adjoint master equation as an alternative approach for the HPZ equation for solving the coefficients in a much simpler form \cite{Carlesso}. The decoherence of quantum Brownian motion results in a slower rate of loss of coherence when the non-Markovianity is weak. Strong non-Markovian nature give a faster rate of decoherence than the Markovian case \cite{Horhammer}. Further, a few studies have been done for exploring  decoherence in two-dimensional Brownian motion models in a non-commutative space, which can be realized by applying a magnetic field in a  direction perpendicular to the two-dimensional plane \cite{Ghorashi}. In the corresponding study linear entropy has been considered as the measure of the purity of the states in non-commutative space \cite{Ghorashi}. The density operator and the Wigner function for an an-harmonic oscillator were computed in a 2D non-commutative phase space and the effects of magnetic field on the decoherence rate were highlighted, using time-independent coefficients, featuring the Markovian heat bath dynamics \cite{Tcoffo, Germain, Armel}. Later, the master equation was derived for the momentum coupling scenario in the framework of non-Markovian dynamics \cite{Ferialdi}. Here we have gone beyond these studies and explored the decoherence of a charged particle in a magnetic field in contact with an environment, with different types of bath spectral densities, suitably generalizing to non-linear densities as well (super-Ohmic and sub-Ohmic). In our work, we have considered the non-Markovian form of the HPZ master equation and gone beyond the Caldeira-Leggett model in deriving  decoherence at both the high-temperature classical regimes and the very low-temperature quantum regime, where the noise kernel for the Ohmic bath is itself non-trivial in nature \cite{Caldeira1,Caldeira2}.

The paper is organised as follows. In Sec. $II$ we derive the master equation for a charged particle in a magnetic field. In Sec. $III$ we consider the solution of the master equation for an Ohmic bath for three different models for the upper cut-off frequency for the spectral density of the bath. We discuss decoherence both in the high-temperature classical regime dominated by thermal fluctuations and in the very low-temperature quantum regime dominated by zero-point fluctuations. We discuss decoherence in the context of sub-Ohmic and super-Ohmic baths in Sec. $IV$. We analyse and discuss our results in detail and present plots of various cases studied in our paper in Sec. $V$ and finally end the paper with a few concluding remarks in Sec. $VI$. An appendix is added at the end for the detailed derivation of the master equation for the corresponding Brownian oscillator model.

\section{master equation for a charged particle in a magnetic field}
We first set up the Born-Markov Master(BMM) equation for the quantum Brownian motion(QBM) of a charged particle linearly coupled to the bath via position coordinates and placed in a magnetic field and then generalise it to non-Markovian models. 

\subsection{Born-Markov Master Equation}
For the derivation of the Born-Markov master equation, the following two assumptions are made\cite{Maximilianbook,Breuer}:\\
(i) \textbf{Born approximation}:- The coupling between the system and the environment is sufficiently weak and the environment sufficiently large so that the effect of the system-environment interaction on the density operator of the environment can be neglected and the system-environment composite density matrix can be taken as a tensor product of the system density matrix and the environmental density matrix, which is taken to be in a stationary state (constant with respect to time).
\begin{align}
    \rho_{SE}(t)\approx \rho_s(t) \otimes \rho_E
\end{align}
(ii) \textbf{Markovian approximation}:- The environment is memory-free and the time evolution equations are time-local.\\ 
The Liouville-Von Neumann equation for the total density operator in the interaction picture is given by \cite{Maximilianbook,Breuer}:
\begin{align}
    \frac{\partial}{\partial t} \rho^{(I)}(t)= \frac{1}{\hbar}\left[H_{int}(t),\rho^{(I)}(t) \right] \label{Neumann}
\end{align}
Taking the trace over the environment and re-writing the interaction Hamiltonian as a diagonal decomposition:
\begin{align}
    H_{int}(t)=\sum_{\alpha}S_{\alpha}(t)\otimes E_{\alpha}(t)  \label{inthamiltonian}
\end{align}
we find the time evolution of the system density matrix:
\begin{widetext}
\begin{align}
    \frac{\partial}{\partial t}\rho_s^{(I)}(t)=-\frac{1}{\hbar^2}\int_0^t dt'\sum_{\alpha \beta} \bigg \lbrace C_{\alpha \beta} (t-t')\left[S_{\alpha}(t) S_{\beta}(t')\rho_s^{(I)}(t')-S_{\beta}(t')\rho_s^{(I)}(t')S_{\alpha}(t) \right]+ \notag \\
  C_{\beta \alpha} (t'-t)\left[ \rho_s^{(I)}(t')S_{\beta}(t')S_{\alpha}(t)-S_{\alpha}(t)\rho_s^{(I)}(t')S_{\beta}(t') \right]  \bigg \rbrace \label{timeevolution}
\end{align}
\end{widetext}
where
\begin{align}
    C_{\alpha \beta}(t-t')=Tr_E \left \lbrace E_{\alpha} (t-t') E_{\beta} \rho_E \right \rbrace=\langle E_{\alpha}(t-t') E_\beta \rangle_{\rho_E} \label{envcorrelation}
\end{align}
$C_{\alpha \beta}(t-t')$ and $C_{\beta\alpha }(t'-t)$ are called the environmental self-correlation functions \cite{Schlosshauer} and $S_\alpha$ and $E_\alpha$ are the system operators and environment operators respectively.\\
For deriving the Markovian form of the master equation, one can replace $t'\rightarrow t-\tau$ and the retarded density matrix   $\rho_s(t')$ by  $\rho_s(t)$, assuming that the change in density matrix is small within the time scale of interest and rewrite Eq.(\ref{timeevolution}) as:
\begin{align}
    \frac{\partial}{\partial t}\rho_s^{(I)}(t)=-\frac{1}{\hbar^2}\int_0^\infty d\tau\sum_{\alpha \beta} \bigg \lbrace C_{\alpha \beta} (\tau)\left[S_{\alpha}(t) S_{\beta}(t-\tau)\rho_s^{(I)}(t)-S_{\beta}(t-\tau)\rho_s^{(I)}(t)S_{\alpha}(t) \right]+ \notag \\
  C_{\beta \alpha} (-\tau)\left[ \rho_s^{(I)}(t)S_{\beta}(t-\tau)S_{\alpha}(t)-S_{\alpha}(t)\rho_s^{(I)}(t)S_{\beta}(t-\tau) \right]  \bigg \rbrace \label{timeevolution2}
\end{align}
Then, the final form of the Born-Markov master equation is obtained as \cite{Redfield,Blum}:
\begin{align}
    \frac{\partial}{\partial t} \rho_s(t)=-\frac{i}{\hbar}\left[H_s,\rho_s(t) \right]-\frac{1}{\hbar^2}\left \lbrace \left[S,B\rho_s(t) \right]  + \left[\rho_s(t)C,S \right] \right \rbrace \label{Born-Markovgen}
\end{align}
where,
\begin{align}
    B_\alpha=\int_0^\infty d\tau \sum_\beta C_{\alpha \beta}(\tau) S_\beta ^{(I)}(-\tau) \label{Balphagen}\\
    C_\alpha=\int_0^\infty d\tau \sum_\beta C_{\beta \alpha}(-\tau) S_\beta ^{(I)}(-\tau) \label{Calphagen}
\end{align}
The operators with superscript $(I)$ are in the interaction picture and the remaining operators are Schr\"odinger picture operators and $\tau=t-t'$, which gives a measure of the time through which the environment retains the memory of the interaction with the system.
\subsection{Non-Markovian Master Equation}
The non-Markovian master equation can also be derived for open systems where the system-environment coupling is still weak \cite{Schlosshauer,Caldeira3,Caldeira1,HuPaz,Unruh,Intravaia,Ferialdi}. In the non-Markovian case, the form of the master equation is the same, however, the operators $B_\alpha$ and $C_\alpha$ are time dependent.
\begin{align}
    B_\alpha(t)=\int_0^t d\tau \sum_\beta C_{\alpha \beta}(\tau) S_\beta ^{(I)}(-\tau)\label{Balphanonmarkov}\\
    C_\alpha(t)=\int_0^t d\tau \sum_\beta C_{\beta \alpha}(-\tau) S_\beta ^{(I)}(-\tau) \label{Calphanonmarkov}
\end{align}
It is to be noted that in the non-Markovian derivation of the master equation, the Born Approximation is assumed to hold true (weak coupling limit) \cite{Maximilianbook}.\\
Now, we consider the specific case of quantum Brownian motion of a charged particle in a harmonic oscillator potential in the presence of a magnetic field. The magnetic field is applied along the $z$ axis and the motion of the charged particle is confined in the two-dimensional $x-y$ plane. The Hamiltonian for this system can be written as: 
\begin{equation}
    H = H_{S} + H_{E}+H_{SE}
\end{equation}
with
\begin{align}
    H_{S} &= \frac{1}{2m}\left(p-\frac{eA}{c}\right)^{2} +
    \frac{1}{2} m\omega_0^{2}(x^{2}+y^{2}) \label{syshamiltonian}\\
     H_{E}&=\sum_{i} \frac{p_{i}^{2}}{2m_{i}} +\frac{1}{2}m_{i}\omega_{i}^{2}q_{i}^{2} \label{envhamiltonian}
\end{align}
where $H_{S}$ and $H_{E}$ are respectively the Hamiltonian of the system and the environment.  $A$ is the vector potential of the applied magnetic field and $p$, $x(y)$, $m$, $\omega_0$ are respectively the momentum, position coordinate, mass and frequency of the harmonic oscillator potential. $p_{i}$, $q_{i}$, $m_{i}$ and $\omega_{i}$ are the
momentum, position coordinate, mass and frequency
of the $i^{th}$ bath oscillator. \\
Using Eq.(\ref{inthamiltonian}), we have modelled the particle-bath interaction in the form: 
\begin{equation}
    H_{SE} = x \otimes \sum_{i} c_{i}q_{ix} + y \otimes \sum_{i} c_{i}q_{iy} \label{sys-envhamiltonian}
\end{equation}
where the coupled $x$ and $y$ coordinates are monitored by the environment. The system is  linearly coupled to the environment via position coordinate and $c_{i}$ is the coupling constant. \\
Using the generalized form of the master equation (Eqs.(\ref{Born-Markovgen},\ref{Balphanonmarkov},\ref{Calphanonmarkov}) and Eqs.(\ref{syshamiltonian},\ref{envhamiltonian},\ref{sys-envhamiltonian})) for the quantum Brownian particle in the presence of a magnetic field, one gets:
\begin{align}
    \frac{\partial \rho_s (t)}{\partial t}=-\frac{i}{\hbar}\left[H_s, \rho_s(t) \right]-&\frac{1}{\hbar}\int_0^t d\tau \bigg \lbrace \nu(\tau)\left[ x,[x(-\tau),\rho_s(t)\right]]-i \eta(\tau))\left[ x,[x(-\tau),\rho_s(t)\right]]- \notag \\
    &  \nu(\tau)\left[ y,[y(-\tau),\rho_s(t)\right]]-i \eta(\tau))\left[ y,[y(-\tau),\rho_s(t)\right]] \bigg \rbrace \label{mastereqBrownian}
\end{align}
where, $\nu(\tau)$ and $\eta(t)$ are the noise and dissipation kernels respectively given by \cite{Maximilianbook,Breuer}:
\begin{align}
   & \nu(\tau)=\int_0^\infty d\omega J(\omega)coth\left( \frac{\omega}{\Omega_{th}}\right) cos(\omega \tau) \label{noisekernel}\\
   & \eta(\tau)=\int_0^\infty d\omega J(\omega) sin(\omega \tau) \label{dissipationkernel}
\end{align}
where, $\Omega_{th}=\frac{2 k_B T}{\hbar}$. \\ \\
$J(\omega)$ is the spectral density of the environment oscillators:
\begin{align}
    J(\omega)=\sum_i \frac{c_i}{2m_i \omega_i}\delta(\omega-\omega_i)
\end{align}
Now, we solve the Heisenberg equations of motion for the system Hamiltonian (Eq.(\ref{syshamiltonian})) and we get the particle position coordinates as:
\begin{align}
    x(\tau)=&\frac{1}{2\omega_c \sqrt{4 \omega_0^2+\omega_c^2}}\bigg[\bigg \lbrace\left(-\omega_c^2+\omega_c \sqrt{4\omega_0^2+\omega_c^2} \right) cosh(A\tau)+\left(\omega_c^2+\omega_c \sqrt{4\omega_0^2+\omega_c^2} \right) cosh(A\tau) \bigg \rbrace X \notag \\
   &\bigg \lbrace 2\sqrt{2} \omega_0^2 \omega_c \left(\frac{sinh(A \tau)}{A}-\frac{sinh(B \tau)}{B} \right) \bigg \rbrace Y+\bigg \lbrace \frac{\sqrt{2}\left(\omega_c^2+\omega_c \sqrt{4\omega_0^2+\omega_c^2} \right)}{A}sinh(A \tau)+ \notag \\ &\frac{\sqrt{2}\left(\omega_c^2+\omega_c \sqrt{4\omega_0^2+\omega_c^2} \right)}{A} sinh(B \tau) \bigg \rbrace V_x +\bigg \lbrace 2\omega_c \left( -cosh(A \tau)+cosh(B \tau)\right) \bigg \rbrace V_y\bigg] \label{couplx}\\
   y(\tau)=&\frac{1}{2\omega_c \sqrt{4 \omega_0^2+\omega_c^2}}\bigg[\bigg \lbrace\left(-\omega_c^2+\omega_c \sqrt{4\omega_0^2+\omega_c^2} \right) cosh(A \tau)+\left(\omega_c^2+\omega_c \sqrt{4\omega_0^2+\omega_c^2} \right) cosh(A \tau) \bigg \rbrace Y  \notag \\
   &\bigg \lbrace 2\sqrt{2} \omega_0^2 \omega_c \left(\frac{sinh(A \tau)}{A}-\frac{sinh(B \tau)}{B} \right) \bigg \rbrace X+\bigg \lbrace \frac{\sqrt{2}\left(\omega_c^2+\omega_c \sqrt{4\omega_0^2+\omega_c^2} \right)}{A}sinh(A \tau)+ \notag \\ &\frac{\sqrt{2}\left(\omega_c^2+\omega_c \sqrt{4\omega_0^2+\omega_c^2} \right)}{A} sinh(B \tau) \bigg \rbrace V_y -\bigg \lbrace 2\omega_c \left( -cosh(A \tau)+cosh(B \tau)\right) \bigg \rbrace V_x\bigg] \label{couply}
   \end{align}
   where $\omega_c=eB/m$ is the cyclotron frequency and
   \begin{align}
       &A=\frac{\sqrt{-2\omega_0^2-\omega_c^2-\omega_c\sqrt{4\omega_0^2+\omega_c^2}}}{2} \\
      & B=\frac{\sqrt{-2\omega_0^2-\omega_c^2+\omega_c\sqrt{4\omega_0^2+\omega_c^2}}}{2}\\
       &X=x(0), Y=y(0), V_x=P_x/m=\Dot{x}(0), V_y=P_y/m=\Dot{y}(0)
   \end{align}
    $X$ and $Y$ are the position operators in Schr\"{o}dinger picture and
   $V_x$, $V_y$, $P_x$, $P_y$ are the initial velocities and momentum in x and y direction respectively, representing the Schr\"{o}dinger picture velocity and momentum operators.\\
   Using Eq.(\ref{mastereqBrownian}) and retaining only  
   the decoherence terms we get (the full expression can be found in the Appendix):
   \begin{align}
       \frac{\partial \rho_s}{\partial t}=-\frac{1}{\hbar}\int_0^td\tau \nu(\tau)F_1(\tau)\left[X,\left[X,\rho_s(t) \right]  \right]-\frac{1}{\hbar}\int_0^td\tau \nu(\tau)F_1(\tau)\left[Y,\left[Y,\rho_s(t) \right]  \right]-\notag \\
       \frac{1}{\hbar}\int_0^td\tau \nu(\tau)F_2(\tau)\left[X,\left[Y,\rho_s(t) \right]  \right]-\frac{1}{\hbar}\int_0^td\tau \nu(\tau)F_2(\tau)\left[Y,\left[X,\rho_s(t) \right]  \right] \label{decequation}
   \end{align}
   where, 
   \begin{align}
       F_1(\tau)=&\frac{\left(-\omega_c+\sqrt{4\omega_0^2+\omega_c^2} \right)cosh(A \tau)+\left(\omega_c+\sqrt{4\omega_0^2+\omega_c^2} \right)cosh(B \tau)}{2\sqrt{4\omega_0^2+\omega_c^2}} \label{F1}\\
   &F_2(\tau)=\frac{2\sqrt{2}\omega_0^2\left(-B sinh(A \tau)+Asinh(B \tau) \right)}{2AB\sqrt{4\omega_0^2+\omega_c^2}} \label{F2}
   \end{align}
   
In Eq.(\ref{decequation}), 
one notices that the terms on the right 
hand side pertains to spatial decoherence. There are cross terms between $X$ and $Y$, which signify that the decoherence in $x$ and $y$ directions are interlinked.\\
The off-diagonal elements of the density matrix decay as:
\begin{align}
   \rho_s(x,x',y,y',t)=\rho_s(x,x',y,y',0) exp \left[-\int_0^{t}D(t')dt'\right] \label{densitydecay}
\end{align}
where,

\begin{equation}
 D(t)=\lambda_1(t) \left \lbrace \left( \Delta x\right)^2 +
 \left(\Delta y\right)^2 \right \rbrace +2 \lambda_2(t)\left \lbrace\left( \Delta x \right) \left(\Delta y \right) \right \rbrace \label{decorate}
\end{equation}
Here, $\Delta x=x-x^{'}$ and $\Delta y=y-y^{'} $.
\begin{align}
   &\lambda_1(t)=\frac{1}{\hbar}\int_{0} ^{t} d \tau \nu(\tau)F_1 (\tau) \label{lambda1}\\
   & \lambda_2(t)=\frac{1}{\hbar}\int_{0} ^{t} d \tau \nu(\tau)F_2 (\tau) \label{lambda2}
\end{align}
In order to obtain explicit expressions for $\lambda_1$ and $\lambda_2$ we consider an Ohmic spectral density ($J(\omega)\propto \omega$) for the bath.
In this study, we have considered three types of cut-off models for the Ohmic bath, for which the spectral densities are listed below in a tabular form:

\begin{center}
\begin{table}[H]
\caption{ Table for the Ohmic spectral densities with different types of cut-off}
\centering
\begin{tabular}{ |p{4cm}|p{4cm}|p{4cm}| }

 \hline
 \centering{Abrupt cut-off}  & \centering{Drude-Lorentz cut-off} &  \centering{Exponential cut-off}\cr
\hline
  \centering{$J(\omega)=\gamma \omega \Theta(\Lambda-\omega)$} & \centering{$J(\omega)=\gamma \omega \frac{\Lambda^2}{\Lambda^2+\omega^2}$} & \centering{$J(\omega)=\gamma \omega exp \left(-\omega/\Lambda  \right)$}    \cr
 \hline

\end{tabular}

\label{table1}
\end{table}
\end{center}
$\Theta$ in Table(\ref{table1}) represents the Heaviside theta function and $\Lambda$ is the ultraviolet  cut-off. Here, $\gamma$ represents the effective coupling between the system 
 and the environment.\\

\begin{figure}[H]
\centering
\includegraphics[scale=0.65]{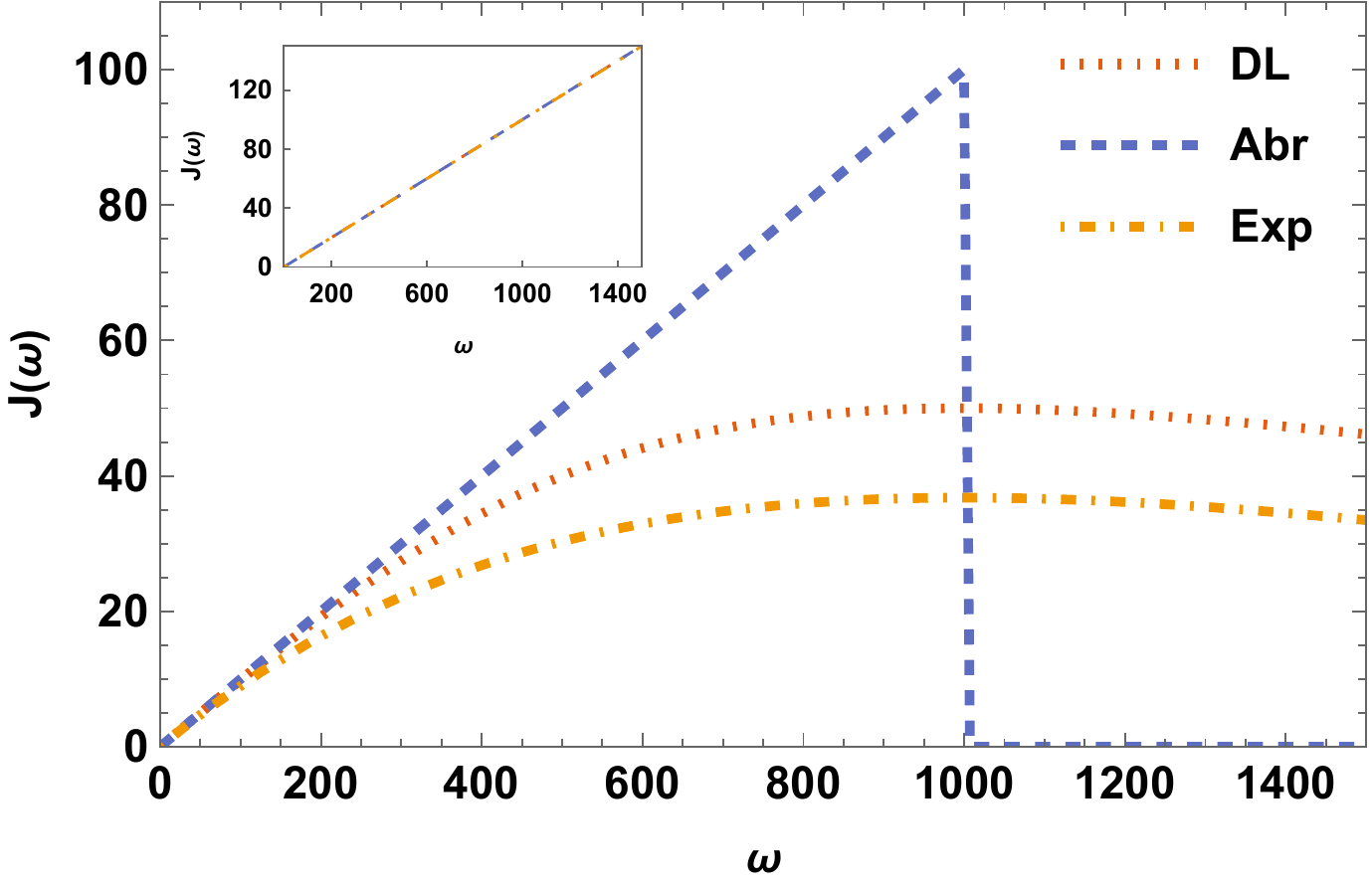}

\caption{Plot showing the three types of spectral density versus $\omega$ for different cut-offs ($\Lambda=10^3$) and the inset plot shows the convergence of those three spectral densities at a very high cut-off ($\Lambda=10^{6})$ value ($J(\omega)$ is in the unit of $\gamma^2/m$ and $\omega$, $\Lambda$ are in the units of $\gamma/m$).}
\label{Fig1}
\end{figure}

\section{Solution for an Ohmic Model}
Here we derive the master equations and study the spatial decoherence for the Ohmic bath model with three different models for the ultra-violet cutoff:
\subsection{Abrupt cut-off}
As already listed in the table, for the abrupt cut-off case, the spectral density grows linearly with $\omega$ up to the cut-off frequency $\Lambda$ and then exhibits a sudden jump to zero for $\omega > \Lambda$:
\begin{equation}
    J(\omega)=
    \begin{cases}
      \gamma \omega, & \text{for}\ \omega<\Lambda \\
      0, & \hspace{0.6cm}\omega>\Lambda
    \end{cases}
  \end{equation}
\subsubsection{\textbf{High-temperature classical domain}}
In the high-temperature classical domain,
\begin{align}
    coth\left( \frac{\omega}{\Omega_{th}} \right) \approx \frac{\Omega_{th}}{\omega} \label{cothhightemp}
\end{align}
Using Eq.(\ref{cothhightemp}) in Eq.(\ref{noisekernel}), one gets:
\begin{align}
    \nu(\tau)=\int_0^\Lambda \gamma \Omega_{th} cos(\omega \tau) d\omega = \frac{\gamma \Omega_{th} sin(\Lambda \tau)}{\tau} \label{nuabrupt}
\end{align}
From the derived $\nu(\tau)$ (Eq.(\ref{nuabrupt})), it is straightforward to compute $\lambda_1(t)$ and $\lambda_2(t)$ using Eq.(\ref{lambda1}) and Eq.(\ref{lambda2}) respectively.
\begin{align}
    &\lambda_1(t)=\frac{\gamma \Omega_{th}}{2 \hbar}\big( -M f_{1}\left(t(A'-\lambda) \right)-P  f_{1}\left(t(B'-\lambda) \right)+ M f_{1}\left(t(A'+\lambda) \right)+ P  f_{1}\left(t(B'+\lambda) \right)\big) \label{lambda1abrupt}\\
    &\lambda_2(t)=\frac{-i\Omega_{th} \gamma}{2\hbar}G \bigg( B \left \lbrace  f_{2}\left(t(\Lambda-A')\right)  - f_{2}\left(t(\Lambda+A')\right)+log\left(\frac{\Lambda+A'}{\Lambda-A'}\right)\right \rbrace -\notag \\
    &A \left \lbrace f_{2}\left(t(\Lambda-B')\right)  -f_{2}\left(t(\Lambda+B')\right)+log\left(\frac{\Lambda+B'}{\Lambda-B'}\right)\right \rbrace \bigg) \label{lambda2abrupt}
\end{align}
where, $A=iA', B=iB'$ and
\begin{align}
  M=\frac{-\omega_c+\sqrt{4\omega_0^2+\omega_c^2}}{2\sqrt{4\omega_0^2+\omega_c^2}},
    P=\frac{\omega_c+\sqrt{4\omega_0^2+\omega_c^2}}{2\sqrt{4\omega_0^2+\omega_c^2}},
    G=\frac{2\sqrt{2} \omega_0^2}{2 \sqrt{4\omega_0^2+\omega_c^2}} \label{M,P}
    \end{align}
    
    \begin{align}
    f_{1}(z)=sinIntegral[z]=\int_0^z\frac{sint}{t}dt \\
     f_{2}(z)= cosIntegral[z]=\int_0^z\frac{cost}{t}dt 
     \end{align}

Inserting the expressions for $\lambda_1$ and $\lambda_2$ in Eq.(\ref{densitydecay}) we get:
\begin{align}
    \frac{\rho_s(t)}{\rho_s(0)} =  exp\left[-\left(D_1(t)+D_2(t)\right)\right] \label{rho/rho0}
\end{align}
where,
\begin{align}
   & D_1(t)=\left \lbrace \left( \Delta x \right)^2+\left(\Delta y  \right)^2  \right \rbrace\int_0^t \lambda_1(t')dt' \label{D1(t)}\\
   & D_2(t)=\left \lbrace \left( \Delta x \right)\left( \Delta y \right)  \right \rbrace \int_0^t \lambda_2(t')dt' \label{D2(t)}
\end{align}
Using Eqs.(\ref{D1(t)}) and (\ref{D2(t)}):
\begin{align}
    D_1(t)&=\frac{\gamma \Omega_{th}}{2 \hbar} \left \lbrace\left( \Delta x \right)^2+ \left(  \Delta y \right)^2  \right \rbrace \bigg( \frac{2M \left[ \Lambda \left(-1+cos(\Lambda t)cos(A't)\right)+sin(\Lambda t) sin(A't)  \right]}{\Lambda^2-(A')^2} + \notag \\
    &\frac{P \left[ -1+cos\left((\Lambda-B')t \right) \right]}{\Lambda-B'}+\frac{P \left[ -1+cos\left((\Lambda+B')t \right) \right]}{\Lambda
    +B'}+\big[ M\left[f_1\left((\Lambda-A')t  \right)+f_1\left((\Lambda+A')t  \right)  \right]+ \notag \\
  &  P\left[f_1\left((\Lambda-B')t  \right)+f_1\left((\Lambda+B')t  \right)  \right] \big]t \bigg) \label{D1expression}
\end{align}

\begin{align}
    D_2(t)=&\left(\frac{i\Omega_{th}\gamma G}{2 \hbar}\right) \left \lbrace \left(\Delta x \right) \left( \Delta y \right)\right \rbrace \frac{1}{t} \bigg( \frac{Bt sin \left((\Lambda-A')t \right)}{\Lambda-A'}    -\frac{Bt sin \left((\Lambda+A')t \right)}{\Lambda+A'}  -\frac{At sin \left((\Lambda-B')t \right)}{\Lambda-B'} +\notag\\
    &\frac{At sin \left((\Lambda+B')t \right)}{\Lambda+B'} +\bigg\lbrace -B \left[ f_2\left( (\Lambda-A')t \right)- f_2\left( (\Lambda+A')t \right) \right] +A \big[f_2\left( (\Lambda-B')t \right)- \notag \\
   & f_2\left( (\Lambda+B')t \right) \big]+ 
     B \hspace{0.1cm} log\left( \frac{\Lambda-A'}{\Lambda+A'} \right)-A \hspace{0.1cm} log\left( \frac{\Lambda-B'}{\Lambda+B'} \right) \bigg \rbrace t^2 \bigg) \label{D2expression}
\end{align}
\textbf{Decoherence at long times:\\\\}
At long times, for $t\rightarrow \infty$ we get:
\begin{align}
    f_1\left( (\Lambda \pm A') \right)=f_1\left( (\Lambda \pm B') \right)\approx\pm \frac{\pi}{2} \label{sin/integrallongtime}
\end{align}

Furthermore, the oscillatory sine and cosine terms do not contribute to the decay at long times and considering the ultraviolet cutoff $\Lambda$ to be very large ($>>\omega,\omega_c$), one can obtain from Eq.(\ref{D1expression}),
\begin{align}
    D_1(t)&\approx \frac{\gamma \Omega_{th}}{2 \hbar} \left \lbrace \left(\Delta x \right)^2+ \left(\Delta y \right)^2\right \rbrace \left( M+P  \right)t \\
    &\approx  \frac{\gamma \Omega_{th}}{2 \hbar} \left \lbrace \left(\Delta x \right)^2+ \left(\Delta y \right)^2\right \rbrace t \label{D1abrupt}
\end{align}
where, $M+P=1$ (see Eq.(\ref{M,P})). \\
Again, at long times,
\begin{align}
    f_2\left( (\Lambda \pm A') \right)=f_2\left( (\Lambda \pm B') \right)\approx0 \label{cosIntegrallabel}
\end{align}
 From Eq.(\ref{D2expression}), one notices, as in the case of  $D_1(t)$, the sine terms of $D_2(t)$ do not contribute to the decay of the off-diagonal terms. \\
 Therefore,
 \begin{align}
     D_2(t)\approx \left( \frac{i \Omega_{th} \gamma G}{2 \hbar} \right)\left \lbrace \left( \Delta x \right) \left( \Delta y \right) \right \rbrace \left[ B \hspace{0.1cm} log\left( \frac{\Lambda-A'}{\Lambda+A'} \right)-A \hspace{0.1cm}log\left( \frac{\Lambda-B'}{\Lambda+B'} \right)\right] t \label{D2abrupt}
 \end{align}
 $D_2(t)$ being imaginary, 
 provides a coherent phase factor $\phi$ and the decay of coherence is determined by $D_1(t)$ (Eq.(\ref{rho/rho0})):
 \begin{align}
     \frac{\rho_s(t)}{\rho_s(0)}=\phi 
 \exp\left[- \frac{\gamma \Omega_{th}}{2 \hbar} \left \lbrace \left(\Delta x \right)^2+ \left(\Delta y \right)^2\right \rbrace t \right]
 \end{align}
 As in the case of a neutral classical Brownian particle, we notice that at high temperatures\cite{Caldeira1,Caldeira2} and long times the off-diagonal elements of the density matrix exhibit an exponential decay and the decoherence time is independent of the cyclotron frequency ($\omega_c$). 
 The cyclotron frequency and the Brownian particle harmonic oscillator frequency only contribute to the coherent part of the off-diagonal elements of the reduced density matrix. The coherent oscillations get suppressed by decoherence due to coupling to the bath. 
 \subsubsection{\textbf{Low-temperature quantum domain}}
 At low temperatures, $\Omega_{th}<<\omega$,
 \begin{align}
     coth\left( \frac{\omega}{\Omega_{th}}\right) \approx 1 \label{cothlowtemp}
 \end{align}
 Then, using Eq.(\ref{noisekernel}),
 \begin{align}
     \nu(\tau)= \gamma \left[\frac{-1+cos(\Lambda \tau)+\Lambda \tau sin(\Lambda \tau)}{\tau^2} \right] \label{nulowtempabrupt}
 \end{align}
 As in the classical case, one can calculate $\lambda_1$ and $\lambda_2$ using Eq.(\ref{nulowtempabrupt}) and Eqs.(\ref{lambda1},\ref{lambda2}):
 \begin{align}
\lambda_1(t)=\frac{\gamma}{\hbar}\bigg[\frac{1}{2t}&\bigg \lbrace  -2\left( -1+cos(\Lambda t)   \right)\left(M cos(A't)+ Pcos(B't)  \right) +MA' t \big[ f_1\left( (\Lambda-A')t \right) +2 f_1\left( A't  \right) -\notag \\
&f_1\left((\Lambda+A')t \right) \big]+ PB' t \left[ f_1\left( (\Lambda-B')t \right) +2 f_1\left( B't  \right) -f_1\left((\Lambda+B')t \right) \right] \bigg \rbrace   \bigg] \label{lambda1lowtempabrupt}
\end{align}
\begin{align}
    \lambda_2(t)= &-\frac{i \gamma G}{8 \hbar} \bigg[\frac{8  \left(-1+cos(\Lambda t) \right) \left(-B sin(A' t)+A sin (B't)  \right)}{t} +2BA'\bigg \lbrace f_3\big( -i(\Lambda-A')t \big)+\notag \\
   &f_3\left( i(\Lambda-A')t \right) -  2f_3\left( -iA't \right)-2f_3\left( iA't \right)+f_3\left( -i(\Lambda+A')t \right) +f_3\left( i(\Lambda+A')t \right)\bigg \rbrace-\notag \\
   &2AB' \bigg \lbrace f_3\left( -i(\Lambda-B')t \right)+f_3\left( i(\Lambda-B')t \right) -  2f_3\left( -iB't \right)-2f_3\left( iB't \right)+f_3\left( -i(\Lambda+B')t \right) +\notag \\
   &f_3\left( i(\Lambda+B')t \right)   \bigg \rbrace \bigg] \label{lambda2lowtempabrupt}
\end{align}
\textbf{Decoherence at long times:\\\\}
As in the classical case, at long times, one can use the approximations for the sinIntegral and cosIntegral functions as used in Eqs.(\ref{sin/integrallongtime}) and (\ref{cosIntegrallabel}). Further, we assume the harmonic oscillator frequency ($\omega_0$) and the cyclotron frequency ($\omega_c$) are small. Thus in the infrared limit \cite{Sinhadecoherence}, $\omega_0 \sim \omega_c << t^{-1}$ and one gets:
\begin{align}
    cos(A't) \approx cos(B't) \approx 1 \\
    f_1 \left(A't \right) \approx f_1 \left( B't \right) \approx 0
\end{align}

Using these and Eqs.(\ref{lambda1}) and (\ref{lambda2}), one can derive the argument of the exponential factor in Eq.(\ref{rho/rho0}):

\begin{align}
    D_1(t)=\frac{\gamma}{\hbar} \left \lbrace \left(\Delta x  \right)^2+\left(\Delta y   \right)^2  \right \rbrace log \left(c t\right)
\end{align}
and \begin{align}
    D_2(t) \approx 0
\end{align}
where,
\begin{align}
    log(c)=&\frac{2}{(\Lambda^2-A'^2)(\Lambda^2-B'^2)}\bigg[\Lambda^4 \left(\Gamma+log(\Lambda)  \right) -\Lambda^2\left( P+\Gamma+ log(\Lambda) \right)B'^2+\notag \\
    &A'^2 \left(-\Lambda^2\left( M+\Gamma+log(\Lambda)  \right)  +1 +\Gamma+log(\Lambda)\right)B'^2\bigg]
\end{align}
$\Gamma$ is the Euler constant $\approx  0.577216$. \\
Therefore, from Eq.(\ref{rho/rho0}):
\begin{align}
\frac{\rho_s(t)}{\rho_s(0)}=(ct)^{-\frac{\gamma}{\hbar}\left \lbrace \left( \Delta x\right)^2+\left( \Delta y\right)^2  \right \rbrace}
\end{align}
We observe that the off-diagonal elements of the density matrix exhibit a power law decay at long times in the low-temperature quantum domain when the harmonic and cyclotron frequencies are in the infrared region. Such a power law trend has been theoretically obtained\cite{Sinhadecoherence} and experimentally observed\cite{Fischer, Hanson, Hanson2,Sarkar} for a neutral  Brownian particle coupled to an Ohmic heat bath with an abrupt ultraviolet cut-off.
\subsection{Drude-Lorentz cut-off}
We have already defined the Drude-Lorentz(DL) spectral density in  Table(\ref{table1}). The analytical form of the DL spectral density is the product of a term proportional to the frequency and a Lorentzian functional form with a cut-off frequency $\Lambda$.  

 Therefore the noise kernel Eq.(\ref{noisekernel}) appears as
\begin{align}
    \nu(\tau)=\frac{\gamma \Lambda^{2}}{2} \int_{-\infty}^{+\infty}  \frac{\omega }{\Lambda^{2}+\omega^{2}} coth \left(\frac{\omega}{\Omega_{th}}\right) cos(\omega \tau) d\omega
    \label{nuLD}
\end{align}
The integrand in the above integral has poles at $\omega = \pm i \Lambda$ and the other poles come from $coth(\frac{\omega}{\Omega_{th}})$ as $\pm i n \pi \Omega_{th}$ ; with $n={1,2,...}$. However, considering only those poles in the lower half of the complex plane we get
\begin{align}
    &\nu(\tau)= \pi\gamma \Lambda^{2}
    \bigg[ \frac{1}{2} cosh(\Lambda\tau)cot\left(\frac{\Lambda}{\Omega_{th}}\right)+ exp(-\pi\Omega_{th}t) \bigg\{\Phi\left(e^{-\pi\Omega_{th}t},1,1-\frac{\Lambda}{\pi\Omega_{th}}\right)\notag\\
    &+\Phi\left(e^{-\pi\Omega_{th}t},1,1+\frac{\Lambda}{\pi\Omega_{th}}\right)\bigg\} +
    exp(\pi\Omega_{th}t) \bigg\{\Phi\left(e^{\pi\Omega_{th}t},1,1-\frac{\Lambda}{\pi\Omega_{th}})+ \Phi(e^{\pi\Omega_{th}t},1,1+\frac{\Lambda}{\pi \Omega_{th}}\right)\bigg\}
    \bigg]
\end{align}
where, $\Phi(z,s,a)$ is the Hurwitz-Lerch(HL) function defined as $\Phi(z,s,a)= \sum_{k=0}^{\infty} z^{k}(k+a)^{-s}$ \cite{Lerch}.
\subsubsection{\textbf{High-temperature Classical regime}}
In order to evaluate the noise kernel in the classical regime, we consider $T \to \infty$; $\Lambda/\pi\Omega_{th} \to 0$, since $\Omega_{th} = 2k_{B}T/\hbar$. In this limit the HL function takes the form  $\Phi(e^{\pi\Omega_{th}t},1,1\pm \frac{\Lambda}{\pi \Omega_{th}}) \approx -e^{-\pi\Omega_{th}t} log(1-e^{\pi\Omega_{th}t})\approx -i \pi e^{-\pi \Omega_{th}t}\left(1-i\Omega_{th}t  \right)$ ;  and $\Phi(e^{-\pi\Omega_{th}t},1,1\pm \frac{\Lambda}{\pi \Omega_{th}})\approx \Phi(0,1,1) =1$ . Further using this approximation we arrive at
\begin{align}
    \nu(\tau)= \frac{\pi\gamma \Lambda^{2}}{2} cot\left(\frac{\Lambda}{\Omega_{th}}\right) cosh(\Lambda\tau)-\frac{i\pi\gamma \Lambda^{2}}{2}\left[1-i\pi \Omega_{th}t \right]
\end{align}
Using the above kernel $\nu(\tau)$ one can calculate the coefficients $\lambda_{1}$ and $\lambda_{2}$ :
\begin{align}
   \lambda_{1}(t) = \frac{\gamma \pi}{2\hbar}\bigg[M g_1(A',t)+P g_1(B',t)  \bigg] 
   \end{align}
   \begin{align}
    \lambda_{2}(t) = \frac{i \gamma G \pi}{2 \hbar} \bigg[ g_2(A,B',t) -g_2(B,A',t)\bigg]
\end{align}
   where,
   \begin{align}
       g_1(z,t)=&\Lambda^2 cot \left( \frac{\Lambda}{\Omega_{th}} \right)\left(\frac{\Lambda cos(zt)sinh(\Lambda t)+cos(\Lambda t)sin(z t)z}{\Lambda^2+z^2}  \right)-\nonumber \\
       &\left(\frac{\pi \Omega_{th}\left(-1+cos(z t) \right)+\left( i+\pi \Omega_{th}t \right) z sin(z t)}{z^2}\right)
        \end{align}

\begin{align}
    g_2(z,v',t)=&\Lambda^2 cot\left(\frac{\Lambda}{\Omega_{th}} \right) \left(\frac{z\left( \Lambda sin(v' t) sin(\Lambda t)+\left( 1-cos(v' t)cosh(\Lambda t)  \right)v'\right)}{\Lambda^2+v'^2}  \right)+\notag \\
    &\left(\frac{z\left(-\pi \Omega_{th}sin(v' t)-iv' \left( 1+\left(-1+i \pi \Omega_{th}t \right)cos(v' t) \right)  \right)}{v'^2}\right)
\end{align}
\subsubsection{\textbf{Low-temperature Quantum regime}}
In the low temperature quantum domain, $T\rightarrow0$ and the HL function $\Phi\left(e^{\pm \pi \Omega_{th}t},1, 1\pm \frac{\Lambda}{\pi \Omega_{th}} \right) \approx 0$. \\In this limit, the noise kernel takes the following form:
\begin{align}
    \nu(\tau)=\frac{\gamma \Lambda^2 \pi}{2} cot \left( \frac{\Lambda}{\Omega_{th}}\right)cosh(\Lambda \tau)  \label{nuDrude}
\end{align}
Then $\lambda_1$ and $\lambda_2$ are calculated from Eq.(\ref{nuDrude}) and Eqs.(\ref{lambda1}, \ref{lambda2}):

\begin{align}
    \lambda_1(t)=\frac{\gamma \pi \Lambda^2}{2 \hbar}cot\left( \frac{\Lambda}{\Omega_{th}} \right)\bigg[ M g_3(A',t)+P g_3(B', t) \bigg]
\end{align}
\begin{align}
    \lambda_2(t)=\frac{-i\gamma \pi \Lambda^2 G}{2 \hbar}cot\left( \frac{\Lambda}{\Omega_{th}} \right) \bigg[g_4(A,B')-g_4(B,A') \bigg]
\end{align}
where,
\begin{align}
    g_3(z,t)=\frac{2\Lambda sin(z t)sinh(\Lambda t)z+cos(z t)cosh(\Lambda t)\left(\Lambda^2-z^2 \right)+z^2-\Lambda^2}{\left(\Lambda^2+z^2\right)^2}
\end{align}

\begin{align}
    g_4(z,v')=\frac{z\left( 2 \Lambda cos(v' t)sinh(\Lambda t)v'+cosh(\Lambda t) sin(v' t)\left( v'^2-\Lambda^2 \right)-v' \left(v'^2+\Lambda^2  \right)t \right)}{\left(\Lambda^2+v'^2\right)^2}
\end{align}
Using the expressions for these $\lambda_1$ and $\lambda_2$, we numerically solved for decoherence factors $D_1(t)$ and $D_2(t)$ and plotted $\frac{\rho_s(t)}{\rho_s(0)}$ as a function of time in both the high temperature classical and the low-temperature quantum domains.

\subsection{Exponential cut-off}
The Ohmic model with an exponential cut-off is defined in Table(\ref{table1}) and shown in Fig():
\begin{align}
    J(\omega)=\gamma \omega e^{-\omega/\Lambda}
\end{align}
\subsubsection{\textbf{High-temperature Classical domain}}
In the high-temperature Classical regime, we use the property of $coth(x) $ as described in Eq.(\ref{cothhightemp}). Hence one can derive the noise kernel using Eq.(\ref{noisekernel}) as:
\begin{align}
    \nu(\tau)=\frac{\gamma\Lambda \Omega_{th}}{1+ \Lambda^2 \tau^2}
\end{align}
As in the earlier cases, we calculate $\lambda_1(t)$ and $\lambda_2(t)$ 
to study decoherence:
\begin{align}
    \lambda_1(t)=\frac{\gamma \Omega_{th}}{2 \hbar}\bigg[M g_5(A',t)+Pg_5(B',t) \bigg]
\end{align}
\begin{align}
    \lambda_2(t)=-\frac{i\gamma G \Omega_{th}}{2 \hbar} \bigg[g_6(A,B',t)-g_6(B,A',t)  \bigg]
\end{align}

where,
\begin{align}
   g_5(z,t)=&-i cosh\left(\frac{z}{\Lambda} \right) \left(f_2\left(\frac{(-i+\Lambda t)z}{\Lambda} \right) -f_2\left(\frac{(i+\Lambda t)z}{\Lambda}  \right)  +i \pi\right)-\notag \\
  & sinh\left( \frac{z}{\Lambda}\right)\left( f_1\left( \frac{(-i+\Lambda t) z}{\Lambda} \right)+ f_1\left( \frac{(i+\Lambda t) z}{\Lambda} \right)\right)
\end{align}
\begin{align}
    g_6(z,v',t)=&z\left[f_2\left(\frac{-i v'}{\Lambda}  \right)+f_2\left(\frac{i v'}{\Lambda}\right) -f_2\left(\frac{(-i+\Lambda t) v'}{\Lambda}\right) -f_2\left(\frac{(i+\Lambda t) v'}{\Lambda}  \right)  \right]sinh\left(\frac{v'}{\Lambda} \right)-\notag \\
    & z cosh\left( \frac{v'}{\Lambda} \right) \left[  2f_1\left( \frac{v'}{\Lambda} \right)-i\left(  f_1\left( \frac{(-i+\Lambda t)v'}{\Lambda} \right)-f_1\left( \frac{(i+\Lambda t)v'}{\Lambda} \right)\right)\right]
\end{align}
\subsubsection{\textbf{Low-temperature Quantum domain}}
As in the case of the earlier cutoff models, in the quantum domain we get,\\
\begin{align}
    \nu(\tau)=\frac{\gamma\left( \frac{1}{\Lambda^2}-\tau^2 \right)}{\left( \frac{1}{\Lambda^2}+\tau^2 \right)^2}
\end{align}
Then $\lambda_1(t)$ and $\lambda_2(t)$ take the form:
\begin{align}
\lambda_1=&\frac{1}{2+2t^2 \lambda^2}\left( \frac{\gamma}{\hbar} \right)\bigg[Mg_7(A',t)+Pg_7(B',t)\bigg]
\end{align}
\begin{align}
    \lambda_2=&\frac{1}{2+2t^2 \lambda^2}\left( \frac{\gamma G}{\hbar} \right)\bigg[g_8(A,B',t)-g_8(B,A',t)\bigg]
\end{align}

where,
\begin{align}
    g_7(z,t)=&2 t \Lambda^2 cos(z t)+\left(1+\Lambda^2 t^2  \right)  \bigg \lbrace i\bigg( f_2\left( \frac{(-i+\Lambda t)z}{\Lambda} \right)- f_2\left( \frac{(i+\Lambda t)z}{\Lambda} \right)+i \pi \bigg) sinh \left(\frac{z}{\Lambda} \right)+\notag \\
&cosh \left( \frac{z}{\Lambda}\right) \bigg(f_1\left( \frac{(-i+\Lambda t)z}{\Lambda} \right)+f_1\left( \frac{(i+\Lambda t)z}{\Lambda} \right)  \bigg) \bigg \rbrace A'
\end{align}
\begin{align}
    g_8(z,v',t)=&2 i \Lambda^2 t z sin(v't)+i z cosh\left( \frac{v'}{\Lambda}\right)f_2 \left( \frac{-i v'}{\Lambda} \right)v'+i z cosh \left( \frac{v'}{\Lambda} \right) \bigg[\Lambda^2 t^2 f_2 \left( \frac{-i v'}{\Lambda} \right) +\notag \\
    &(1+\Lambda^2 t^2) \bigg\{ f_2 \left( \frac{i v'}{\Lambda} \right)-f_2 \left( \frac{(-i+\Lambda t) v'}{\Lambda} \right)-f_2 \left( \frac{(i+\Lambda t) v'}{\Lambda} \right) \bigg\}\bigg]v'+\notag \\
    &z \left( 1+\Lambda^2 t^2 \right) sinh \left(\frac{v'}{\Lambda}  \right)\bigg[-2i f_1\left(\frac{v'}{\Lambda}\right)- f_1\left(\frac{(-i+\Lambda t)v'}{\Lambda}\right)+f_1\left(\frac{(i+\Lambda t)v'}{\Lambda}\right) \bigg]v'
\end{align}
As in the Drude-Lorentz cut-off case, $\frac{\rho_s(t)}{\rho_s(0)}$ is numerically solved and plotted graphically. We make a graphical comparison of the decoherence trends for the three different cut-off models and the results are plotted in Section(V).
\section{DECOHERENCE FOR SUPER-OHMIC AND SUB-OHMIC BATHS}
In this section, we have extended our study to anomalous spectral densities where there is a deviation from the linear dependence on 
$\omega$ of the spectral density. Here we have highlighted the super-Ohmic and sub-Ohmic spectral densities that are given by $J(\omega)\propto \omega^s$, where $s$ can take any values with $s>0$. $s=1$ pertains to the Ohmic spectral density, $s>1$ pertains to the super-Ohmic spectral density and $s<1$ pertains to the sub-Ohmic 
spectral density. As in the Ohmic case, we consider three different types of cut-off models for super-Ohmic and sub-Ohmic spectral densities. In our study, we have focused on two specific cases of super-Ohmic and sub-Ohmic spectral densities with $s=3/2$ and $s=1/2$ respectively, which can be generalized to other values of $s$ using a similar formalism.\\
The spectral densities with $s=3/2$ and $s=1/2$, for three different types of cut-off models and their corresponding noise kernels $\nu(t)$ are tabulated below in Table-\ref{table2} and Table-\ref{table3} respectively:

\begin{center}
\begin{table}[H]
\caption{ Table for the noise kernels of a super-Ohmic spectral density ($s=3/2$) with different types of cut-off}
\centering
\begin{tabular}{ |p{2cm}|p{4.4cm}|p{3.9cm}|p{5.6cm}| }

 \hline
 \centering{} & \centering{Abrupt cut-off}  & \centering{Drude-Lorentz cut-off} &  \centering{Exponential cut-off}\cr
\hline
  \centering{\vspace{0.1cm}$J(\omega)$\\} & \centering{\vspace{0.1cm}$\gamma \omega^{\frac{3}{2}} \Theta(\Lambda-\omega)$\\} & \centering{\vspace{0.1cm}$\gamma \omega^{\frac{3}{2}} \Lambda^2\left(\Lambda^2+\omega^2 \right)^{-1}$\\} & \centering{\vspace{0.1cm}$\gamma \omega^{\frac{3}{2}} exp \left(-\omega/\Lambda  \right)$\\}    \cr
\hline
  \centering{$\nu(\tau)$ at high temp.} & \centering{$\left(\frac{2}{3}\gamma \Lambda^{\frac{3}{2}} \Omega_{th}\right) \times$ \\
 $ \left( {}_{\alpha}F_{\beta}\left[\left\{ \frac{3}{4}\right\} ,\left \{\frac{1}{2},\frac{7}{4}  \right\},\frac{1}{4} \tau^2 \Lambda^2 \right]\right)$} & \centering{$\gamma \Omega_{th} \Lambda^{\frac{3}{2}}\left(1+\Lambda^2 \tau^2  \right)^{-\frac{3}{4}}\times$\\
 $cos\left[ \frac{3}{2} tan^{-1}(\Lambda \tau) \right] \Gamma (\frac{3}{2})$} & \centering{$e^{-\Lambda \tau} \pi \Omega_{th} \Lambda^{\frac{3}{2}}\bigg(1+e^{2 \Lambda \tau} Erfc \left[(\Lambda \tau)^{\frac{1}{2}} \right]-$\\
 $Erfi \left[(\Lambda \tau)^{\frac{1}{2}} \right]\bigg)/2\sqrt{2}$}    \cr
 \hline
 \centering{\vspace{0.2cm}$\nu(\tau)$ at low temp.} & \centering{\vspace{0.05cm}$\left(\frac{2}{5}\gamma \Lambda^{\frac{5}{2}} \Omega_{th}\right)\times$ \\
 $ \left( {}_{\alpha}F_{\beta}\left[\left\{ \frac{5}{4}\right\} ,\left \{\frac{1}{2},\frac{9}{4}  \right\},\frac{1}{4} \tau^2 \Lambda^2 \right]\right)$} & \centering{\vspace{0.05cm}$\gamma \Lambda^{\frac{5}{2}}\left(1+\Lambda^2 \tau^2  \right)^{-\frac{5}{4}}\times$\\
 $cos\left[ \frac{5}{2} tan^{-1}(\Lambda \tau) \right] \Gamma (\frac{5}{2})$} & \centering{$\gamma \Lambda^2 \bigg(\frac{2\sqrt{\pi}}{\tau^{1/4}}-2 \pi \Lambda^{\frac{1}{2}} cosh\left( \Lambda \tau\right)   + $\\
 $\Lambda^{\frac{1}{2}}e^{\Lambda \tau}\pi Erf \left[(\Lambda \tau)^{\frac{1}{2}} \right]-$\\
 $\Lambda^{\frac{1}{2}}e^{-\Lambda \tau}\pi Erfi \left[(\Lambda \tau)^{\frac{1}{2}} \right]\bigg)/2\sqrt{2}$}    \cr
 \hline
\end{tabular}

\label{table2}
\end{table}
\end{center}

\begin{center}
\begin{table}[H]
\caption{ Table for the noise kernels of a sub-Ohmic spectral density ($s=1/2$) with different types of cut-off}
\centering
\begin{tabular}{ |p{2cm}|p{4.4cm}|p{3.9cm}|p{5.6cm}| }

 \hline
 \centering{} & \centering{Abrupt cut-off}  & \centering{Drude-Lorentz cut-off} &  \centering{Exponential cut-off}\cr
\hline
  \centering{\vspace{0.1cm}$J(\omega)$\\} & \centering{\vspace{0.1cm}$\gamma \omega^{\frac{1}{2}} \Theta(\Lambda-\omega)$\\} & \centering{\vspace{0.1cm}$\gamma \omega^{\frac{1}{2}} \Lambda^2\left(\Lambda^2+\omega^2 \right)^{-1}$\\} & \centering{\vspace{0.1cm}$\gamma \omega^{\frac{1}{2}} exp \left(-\omega/\Lambda  \right)$\\}    \cr
\hline
  \centering{$\nu(\tau)$ at high temp.} & \centering{$\left(2\gamma \Lambda^{\frac{1}{2}} \Omega_{th}\right) \times$ \\
 $ \left( {}_{\alpha}F_{\beta}\left[\left\{ \frac{1}{4}\right\} ,\left \{\frac{1}{2},\frac{5}{4}  \right\},\frac{1}{4} \tau^2 \Lambda^2 \right]\right)$} & \centering{$\gamma \Omega_{th} \Lambda^{\frac{1}{2}}\left(1+\Lambda^2 \tau^2  \right)^{-\frac{1}{4}}\times$\\
 $cos\left[ \frac{1}{2} tan^{-1}(\Lambda \tau) \right] \Gamma (\frac{1}{2})$} & \centering{$e^{-\Lambda \tau} \pi \Omega_{th} \Lambda^{\frac{1}{2}}\bigg(1+e^{2 \Lambda \tau} Erfc \left[(\Lambda \tau)^{\frac{1}{2}} \right]+$\\
 $Erfi \left[(\Lambda \tau)^{\frac{1}{2}} \right]\bigg)/2\sqrt{2}$}    \cr
 \hline
 \centering{\vspace{0.05cm}$\nu(\tau)$ at low temp.} & \centering{\vspace{0.05cm}$\left(\frac{2}{3}\gamma \Lambda^{\frac{3}{2}} \Omega_{th}\right)\times$ \\
 $ \left( {}_{\alpha}F_{\beta}\left[\left\{ \frac{3}{4}\right\} ,\left \{\frac{1}{2},\frac{7}{4}  \right\},\frac{1}{4} \tau^2 \Lambda^2 \right]\right)$\\} & \centering{\vspace{0.05cm}$\gamma \Lambda^{\frac{3}{2}}\left(1+\Lambda^2 \tau^2  \right)^{-\frac{3}{4}}\times$\\
 $cos\left[ \frac{3}{2} tan^{-1}(\Lambda \tau) \right] \Gamma (\frac{3}{2})$\\} & \centering{$e^{-\Lambda \tau} \pi \gamma \Lambda^{\frac{3}{2}} \bigg( 1+ e^{2 \Lambda \tau}Erfc \left[(\Lambda \tau)^{\frac{1}{2}}  \right]-$\\
 $Erfi \left[(\Lambda \tau)^{\frac{1}{2}}  \right]\bigg)/2\sqrt{2}$\\}    \cr
 \hline
\end{tabular}

\label{table3}
\end{table}
\end{center}
In the above tables, $\Gamma[z]$ is the Gamma function $\left( \int_0^\infty e^{-t}t^{z-1}dt \right)$ and the function $Erfc[z]$ is the complementary error function defined by $Erfc[z]=1-Erf[z]=1-\frac{2}{\sqrt{\pi}}\int_0^z e^{-t^2}dt$ and the imaginary error function is given by $Erf[z]=Erf[iz]/i$\cite{Andrews}.\\ 
${}_{\alpha}F_{\beta}\left[\{a_1,...,a_p \},\{ b_1,...,b_q \},z 
\right]$ is the hypergeometric function defined by $\sum_{k=0}^{\infty}\frac{(a_1)_k...(a_p)_k}{(b_1)_k...(b_q)_k}\frac{z^k}{k!}$, where $(q)_n$ is the Pochhammer symbol\cite{Szeg}:
\begin{align}
    (q)_n=
    \begin{cases}
      1, & \text{for}\ n=0 \\
      q(q+1)...(q+n-1), & \text{for}\hspace{0.2cm}n>0
    \end{cases}
\end{align}\\
In Tables-(\ref{table2}) and (\ref{table3}) we have displayed the noise kernels at high and low temperatures for two specific super-Ohmic and sub-Ohmic cases with $s=3/2$ and $s=1/2$ respectively. Using these noise kernels we numerically compute the decoherence factor (D(t)) and the plots for $\rho_s(t)/\rho_s(0)$ are shown in the following section.
\section{Results and Discussion}

In this section, we have plotted the numerically computed the reduced system density matrix ($\rho_s(t)/\rho_s(0)$) as a function of time involving three different types of cut-off: Drude-Lorentz (DL), Abrupt (Abr) and Exponential (Exp). The results for the Ohmic ($s = 1$), sub-Ohmic ($s =
1/2$) and super-Ohmic ($s = 3/2$) cases have been exhibited for low-temperature (quantum fluctuation dominated) and high-temperature (thermal fluctuation dominated) regimes.

\begin{figure}[H]
\centering
\includegraphics[scale=0.8]{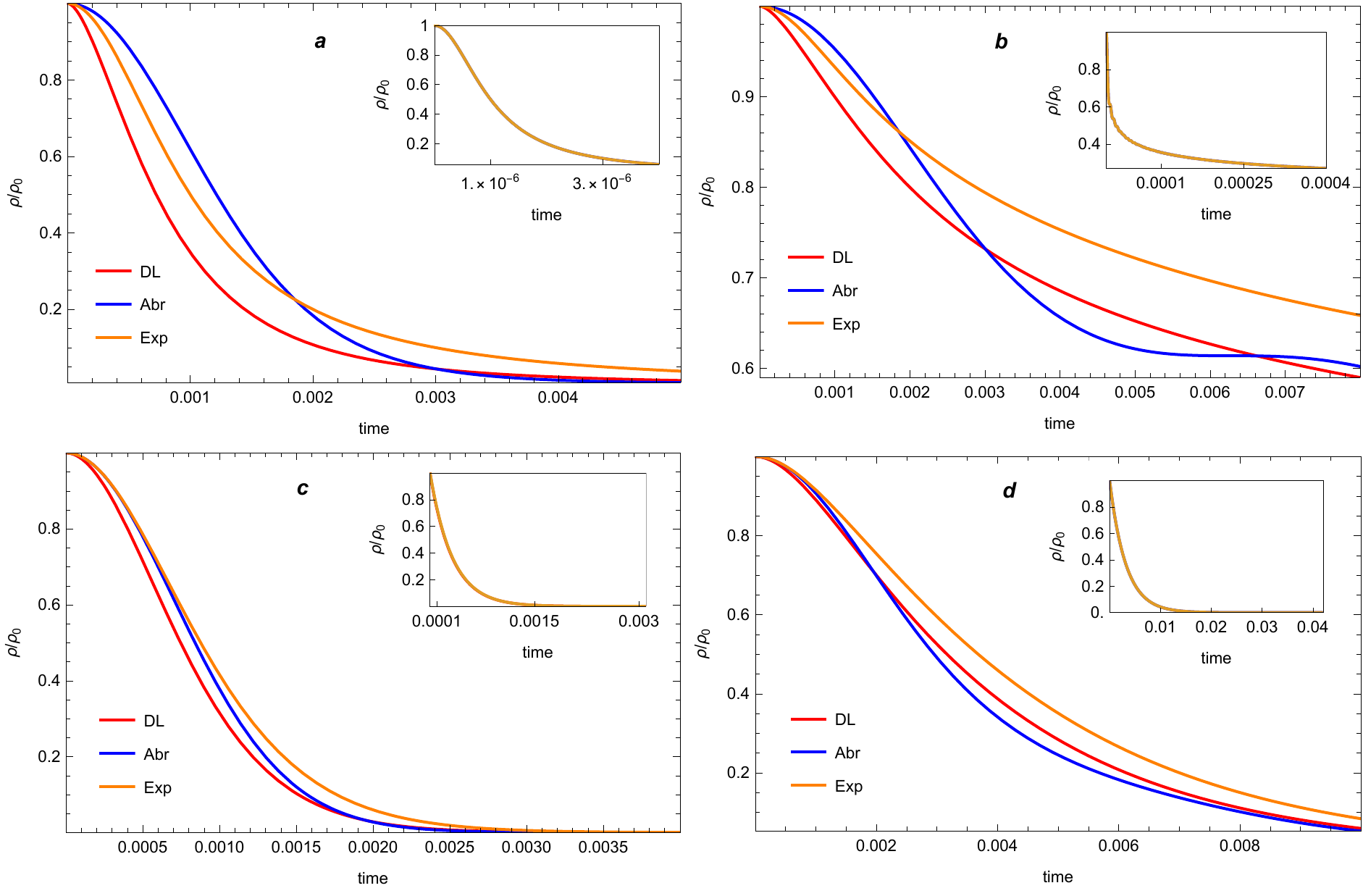}
\caption{$\rho/\rho(0)$ as a function of time for Ohmic (s = 1) bath with cut-off frequency $\Lambda=10^3 \gamma/m$, $\Delta x=\Delta y=1$ \big(units of $\sqrt{\hbar/\gamma}$\big): (a) $\Omega_{th}=0.01$, $\omega_0=10$, $\omega_c=1$, (b) $\Omega_{th}=0.01$, $\omega_0=1$, $\omega_c=10$, (c) $\Omega_{th}=10^3$, $\omega_0=10$, $\omega_c=1$, (d) $\Omega_{th}=10^3$, $\omega_0=1$, $\omega_c=10$. The converged plots for $\Lambda=10^6$ are shown in the insets ($\omega_0$, $\omega_c$, $\Lambda$, $\Omega_{th}$ are in the units of $\gamma/m$ and time is in the unit of $m/\gamma$).  }
\label{Fig2}
\end{figure}

In Fig.(\ref{Fig2}), $\rho_s(t)/\rho_s(0)$ is plotted as a function of time for three different cut-off models at low and high temperatures. One can clearly see from all the plots that the DL curves fall faster than the Exp curves, which can be understood from their spectral density behaviours, where one can see a more flattened curve for the Exp case compared to the DL case (see Fig(\ref{Fig1})). The abrupt cut-off curves for  $\rho_s(t)/\rho_s(0)$ initially show a slower decay rate compared to the other two cut-off models, followed by a faster decay as time is increased (Fig(\ref{Fig2})). This behaviour can be understood from the nature of the spectral density increasing linearly at lower values of frequency $\omega$ and abruptly falling to zero for $\omega> \Lambda$ (Fig.(\ref{Fig1})). Moreover, it is interesting to note that the plots for all models of cut-off converge for very large values of the cut-off frequency, as the spectral densities ($J(\omega)$) also attain convergence and vary linearly with $\omega$, when $\Lambda$ is sufficiently large (Fig.(\ref{Fig1})). 
\begin{figure}[H]
\centering
\hspace{-1cm}\includegraphics[scale=0.952]{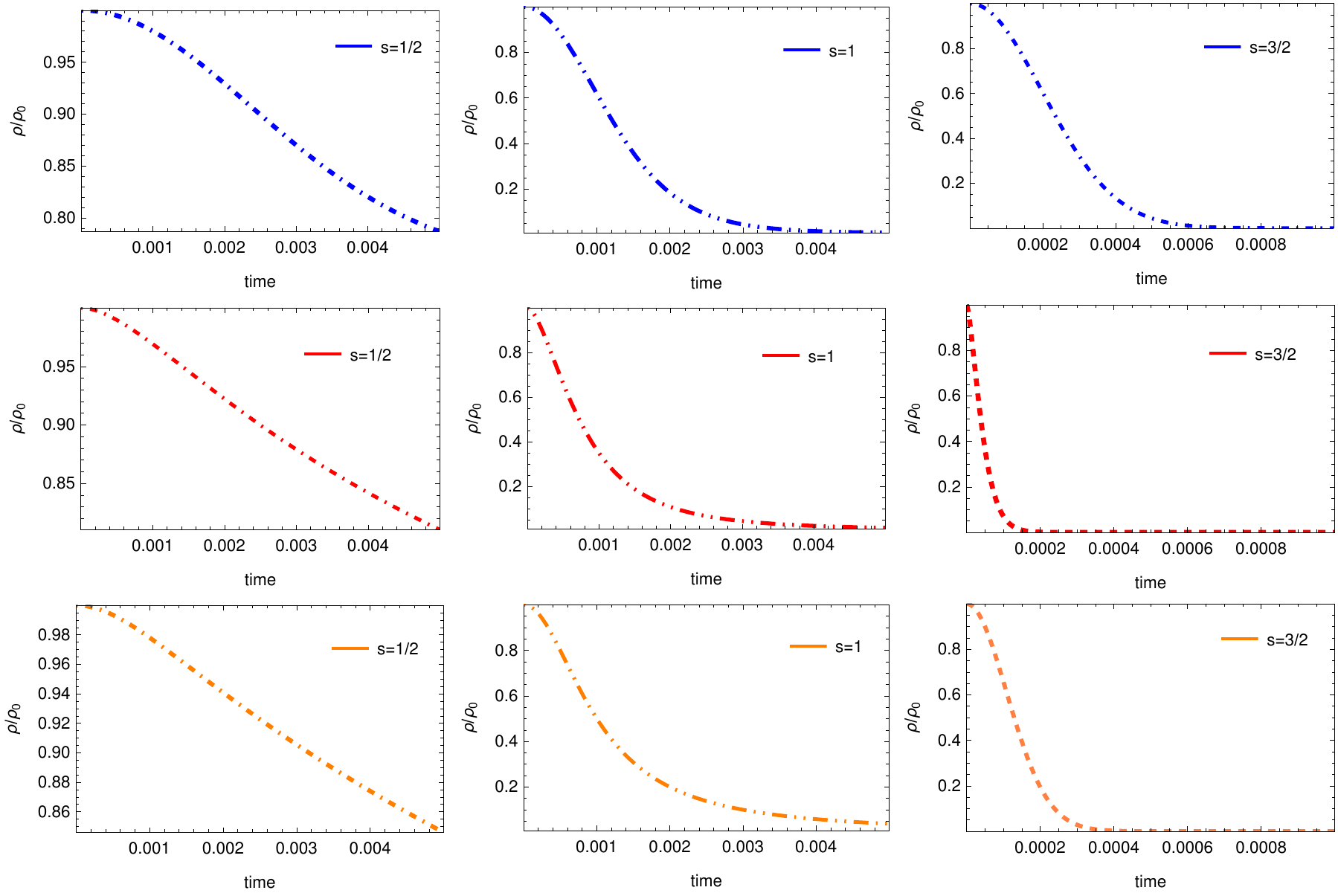}
\caption{$\rho/\rho_0$ for sub-Ohmic ($s=1/2$), Ohmic ($s=1$) and super-Ohmic ($s=3/2$) baths with $\Delta x=\Delta y=1$ \big(units of $\sqrt{\hbar/\gamma}$\big), $\Lambda=10^3$, $\Omega_{th}=0.01$, $\omega_0=10$, $\omega_c=1$. The upper (blue), middle (red) and the lower (orange) panels are for Abrupt, Drude-Lorentz and Exponential cut-offs respectively ($\omega_0$, $\omega_c$, $\Lambda$, $\Omega_{th}$ are in the units of $\gamma/m$ and time is in the unit of $m/\gamma$). }
\label{Fig3}
\end{figure}

\begin{figure}[H]
\centering
\hspace{-1cm}\includegraphics[scale=0.83]{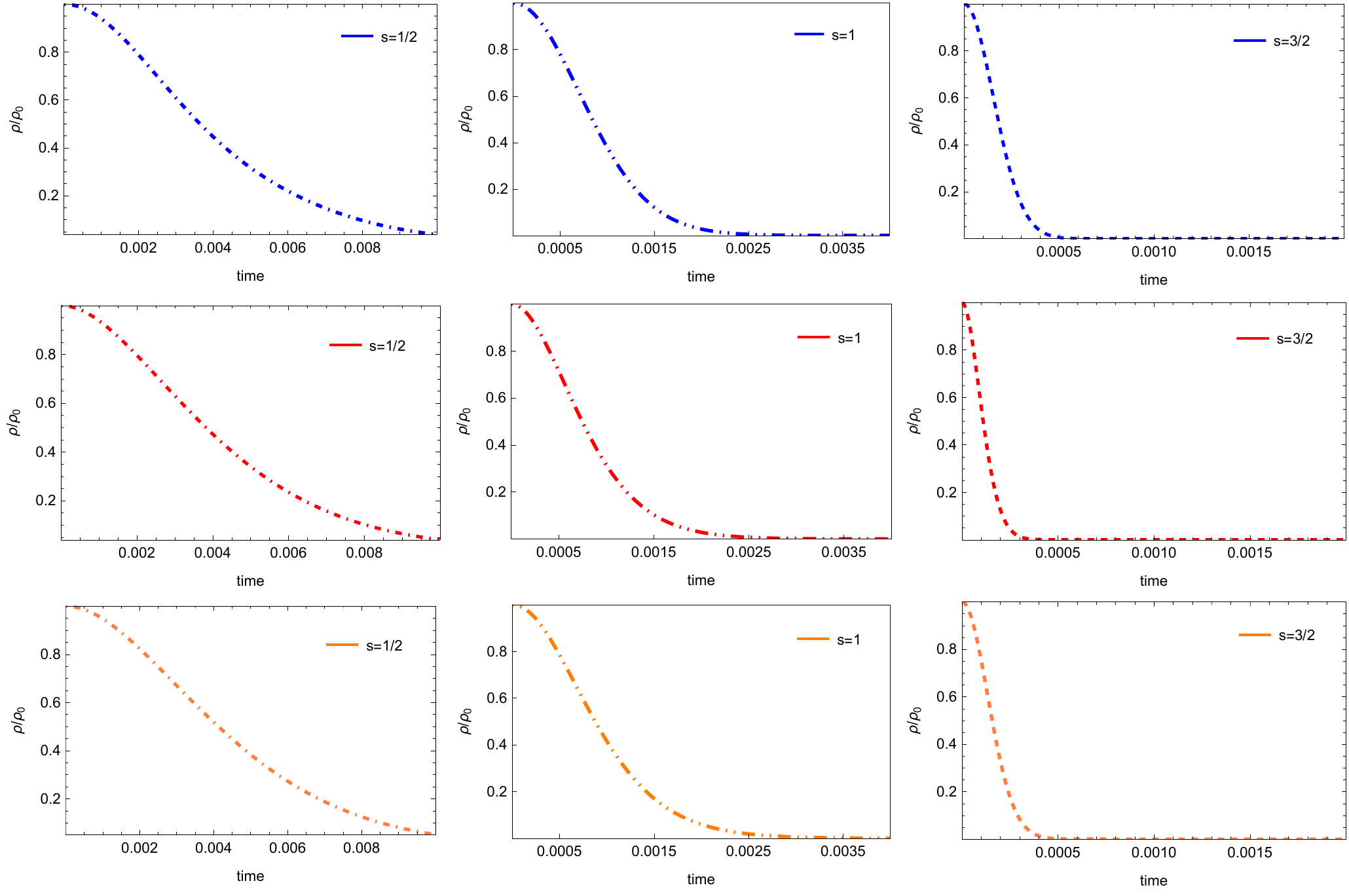}
\caption{$\rho/\rho_0$ for sub-Ohmic ($s=1/2$), Ohmic ($s=1$) and super-Ohmic ($s=3/2$) baths with $\Delta x=\Delta y=1$ \big(units of $\sqrt{\hbar/\gamma}$\big), $\Lambda=10^3$, $\Omega_{th}=10^3$, $\omega_0=10$, $\omega_c=1$. The upper (blue), middle (red) and the lower (orange) panels are for Abrupt, Drude-Lorentz and Exponential cut-offs respectively ($\omega_0$, $\omega_c$, $\Lambda$, $\Omega_{th}$ are in the units of $\gamma/m$ and time is in the unit of $m/\gamma$).}
\label{Fig4}
\end{figure}


The plots for the sub-Ohmic, Ohmic and 
super-Ohmic cases (Figs.(\ref{Fig3}),(\ref{Fig4})) clearly show that  
coherence gets suppressed fastest for a super Ohmic bath and slowest for a sub Ohmic bath with an Ohmic bath exhibiting an intermediate rate of suppression of coherence. Further, it must be noted from the Figs.((\ref{Fig2}),(\ref{Fig3}),(\ref{Fig4})) that the off-diagonal elements of the system density matrix decay at a higher rate for lower values of magnetic field corresponding to all the three bath models and the rate of decoherence is faster for the high-temperature classical regime (Fig.(\ref{Fig4})) in comparison to its quantum counterpart (Fig.(\ref{Fig3})).
\section{Conclusion}
In this paper, we have studied  decoherence in the context of a charged particle in a
magnetic field via a generalised master equation in the high-temperature classical regime dominated by thermal fluctuations and in the low-temperature quantum regime dominated by zero point fluctuations. As in 
the case of a neutral particle, we notice a 
dramatic power law slowing down of decay of coherence in the low-temperature quantum domain at late times ($\omega_0$ $\omega_c$ $<<$ $t^{-1}$ $<<$ $\Lambda$) as opposed to an exponential decay of coherence in the high-temperature classical regime in the limit of long times. This slowing down is a consequence of non-trivial noise correlations in the quantum regime.  

We have studied this for an Ohmic bath, a sub Ohmic bath and a super Ohmic bath for three distinct cutoff models for the baths. As expected,
the results converge in the limit of 
the ultraviolet cutoff going to infinity.
We obtain explicit analytical 
expressions for the temporal decay of 
coherence of the reduced density matrix in the long time limit. In addition, we compare the plots for the various cases based on numerical 
computation. We find that the temporal decay of coherence is fastest for a super Ohmic bath and slowest for a sub Ohmic bath, with an Ohmic bath exhibiting an intermediate rate of suppression of coherence. Again, the rate of decoherence is slower for higher values of the magnetic field, as has been pointed out in previous theoretical and experimental results \cite{Ghorashi,Tcoffo,Germain,Armel}. These results can be later extended to non-commutative space in two dimensions and other anomalous types of couplings can be explored using similar formalism and the outcomes can be further tested in suitable cold atom experiments \cite{Almog,Felinto,Sarkar,Suarez}.
\section{Acknowledgment}
KM acknowledges the hospitality during an academic visit at the Raman Research Institute, Bangalore, where this work was initiated.

\section{Appendix}
The generalized form of the master-equation (Eq.(\ref{mastereqBrownian})) gets the form on substituting the values of $x(-\tau)$ and $y(-\tau)$ as
\begin{align}
    \frac{d\rho_{t}}{dt} = &-\frac{i}{\hbar}\left[H_{s}+\frac{1}{2}m \tilde{\omega}_0^2 \left(X^2+Y^2 \right),\rho_{s}(t)\right]- \frac{1}{\hbar} \int_{0}^{\infty} d\tau \nu(\tau)F_{1}\left[X,\left[X,\rho_{s}(t)\right]\right]\nonumber\\
    &-\frac{1}{\hbar} \int_{0}^{\infty} d\tau \nu(\tau)F_{2}\left[X,\left[Y,\rho_{s}(t)\right]\right]-\frac{1}{\hbar} \int_{0}^{\infty} d\tau \nu(\tau)F_{2}\left[Y,\left[X,\rho_{s}(t)\right]\right] \nonumber\\
   &-\frac{1}{\hbar} \int_{0}^{\infty} d\tau \nu(\tau)F_{1}\left[Y,\left[Y,\rho_{s}(t)\right]\right]
   -\frac{1}{\hbar} \int_{0}^{\infty} d\tau \nu(\tau)F_{4}\left[X,\left[P_{y},\rho_{s}(t)\right]\right]\nonumber\\
   &-\frac{1}{\hbar} \int_{0}^{\infty} d\tau \nu(\tau)F_{3}\left[X,\left[P_{x},\rho_{s}(t)\right]\right] +\frac{1}{\hbar} \int_{0}^{\infty} d\tau \nu(\tau)F_{3}\left[Y,\left[P_{x},\rho_{s}(t)\right]\right]\nonumber\\
   &-\frac{1}{\hbar} \int_{0}^{\infty} d\tau \nu(\tau)F_{4}\left[Y,\left[P_{y},\rho_{s}(t)\right]\right]  +\frac{i}{\hbar} \int_{0}^{\infty} d\tau \eta (\tau) F_{2}\left[X,\{Y
    ,\rho_{s}(t)\}\right] \nonumber\\
   &+ \frac{i}{\hbar} \int_{0}^{\infty} d\tau \eta (\tau) F_{3}\left[X,\{P_{x},\rho_{s}(t)\}\right]+\frac{i}{\hbar} \int_{0}^{\infty} d\tau \eta (\tau) F_{4}\left[X,\{P_{y},\rho_{s}(t)\}\right] \nonumber\\
   &+\frac{i}{\hbar} \int_{0}^{\infty} d\tau \eta (\tau) F_{2}\left[Y,\{X,\rho_{s}(t)\}\right] -\frac{i}{\hbar} \int_{0}^{\infty} d\tau \eta (\tau) F_{3}\left[Y,\{P_{x},\rho_{s}(t)\}\right]\nonumber\\
   &+\frac{i}{\hbar} \int_{0}^{\infty} d\tau \eta (\tau) F_{4}\left[Y,\{P_{y},\rho_{s}(t)\}\right]\label{fulleq}
\end{align}
where
\begin{align}
\tilde{\omega}_0&= -\frac{2}{m} \int_0^{\infty} d\tau \eta(\tau) F_1(\tau)\\
    F_{1}(\tau) &= \frac{\left(-\omega_{c}+\sqrt{4\omega_{0}^{2}+\omega_{c}^{2}}\right)cosh(A \tau)+\left(\omega_{c} + \sqrt{4\omega_{0}^{2}+\omega_{c}^{2}}\right) cosh(B \tau)}{2\omega_{c}\sqrt{4\omega_{0}^{2}+\omega_{c}^{2}}}\\
    F_{2}(\tau) &= \frac{\sqrt{2}\omega_{0}^{2}\left(-sinh(A \tau)/A+sinh(B \tau)/B\right)}{\sqrt{4\omega_{0}^{2}+\omega_{c}^{2}}}\\
    F_{3}(\tau) &= -\frac{\left(\omega_{c} + \sqrt{4\omega_{0}^{2}+\omega_{c}^{2}}\right)sinh(A \tau)/A+\left(-\omega_{c}+\sqrt{4\omega_{0}^{2}+\omega_{c}^{2}}\right)sinh(B \tau)/B}{m\sqrt{2(4\omega_{0}^{2}+\omega_{c}^{2})}}\\
    F_{4}(\tau) &=\frac{2\omega_{c}\left(cosh(A \tau)+cosh(B \tau)\right)}{m\sqrt{4\omega_{0}^{2}+\omega_{c}^{2}}} 
\end{align}
 X,Y, $P_x$, $P_y$, A, B are all defined in the main text.\\
The $\tilde{\omega}_0$ in the first term of Eq.(\ref{fulleq}) represents the frequency shift of the harmonic oscillator, the second to fifth terms containing double commutator form of the two position coordinates give the spatial decoherence, the sixth to ninth terms with one position and one momentum coordinate describe momentum damping and the rest of the terms also represent decoherence, however, their contributions of the anomalous diffusion terms are negligible with respect to the normal decoherence terms \cite{Schlosshauer, Maximilianbook}. 
 So Eq.(\ref{fulleq}) can be written as:
\begin{align}
   \frac{d\rho_{t}}{dt} = -\frac{i}{\hbar}\left[H_{s},\rho_{s}(t)\right] - &\frac{1}{\hbar} \int_{0}^{\infty} d\tau \nu(\tau)F_{1}\left[X,\left[X,\rho_{s}(t)\right]\right] -\frac{1}{\hbar} \int_{0}^{\infty} d\tau \nu(\tau)F_{1}\left[Y,\left[Y,\rho_{s}(t)\right]\right]- \nonumber \\
   & \frac{1}{\hbar} \int_{0}^{\infty} d\tau \nu(\tau)F_{2}\left[X,\left[Y,\rho_{s} (t)\right]\right]
   - \frac{1}{\hbar} \int_{0}^{\infty} d\tau \nu(\tau) F_{2} \left[ Y, \left[ X, \rho_s (t) 
   \right] \right]
\end{align}
Therefore, 
\begin{align}
    \rho_{s}(x, x^{'} ,y, y^{'}, t) = exp \left[ \int_0^t\left(-\lambda_{1}(t)\{(\Delta x)^{2}+(\Delta y)^{2}\}+\lambda_{2} (t)(\Delta x)(\Delta y) \right) dt' \right] \rho_{s}(x,x^{'},y,y^{'},0)
\end{align}

where, $\Delta x = x-x^{'}$ ; $\Delta y = y-y^{'}$. $\lambda_{i}(t)$; ($i=\{1,2\}$) is defined in the main text.

\begin{thebibliography}{51}%
\makeatletter
\providecommand \@ifxundefined [1]{%
 \@ifx{#1\undefined}
}%
\providecommand \@ifnum [1]{%
 \ifnum #1\expandafter \@firstoftwo
 \else \expandafter \@secondoftwo
 \fi
}%
\providecommand \@ifx [1]{%
 \ifx #1\expandafter \@firstoftwo
 \else \expandafter \@secondoftwo
 \fi
}%
\providecommand \natexlab [1]{#1}%
\providecommand \enquote  [1]{``#1''}%
\providecommand \bibnamefont  [1]{#1}%
\providecommand \bibfnamefont [1]{#1}%
\providecommand \citenamefont [1]{#1}%
\providecommand \href@noop [0]{\@secondoftwo}%
\providecommand \href [0]{\begingroup \@sanitize@url \@href}%
\providecommand \@href[1]{\@@startlink{#1}\@@href}%
\providecommand \@@href[1]{\endgroup#1\@@endlink}%
\providecommand \@sanitize@url [0]{\catcode `\\12\catcode `\$12\catcode
  `\&12\catcode `\#12\catcode `\^12\catcode `\_12\catcode `\%12\relax}%
\providecommand \@@startlink[1]{}%
\providecommand \@@endlink[0]{}%
\providecommand \url  [0]{\begingroup\@sanitize@url \@url }%
\providecommand \@url [1]{\endgroup\@href {#1}{\urlprefix }}%
\providecommand \urlprefix  [0]{URL }%
\providecommand \Eprint [0]{\href }%
\providecommand \doibase [0]{https://doi.org/}%
\providecommand \selectlanguage [0]{\@gobble}%
\providecommand \bibinfo  [0]{\@secondoftwo}%
\providecommand \bibfield  [0]{\@secondoftwo}%
\providecommand \translation [1]{[#1]}%
\providecommand \BibitemOpen [0]{}%
\providecommand \bibitemStop [0]{}%
\providecommand \bibitemNoStop [0]{.\EOS\space}%
\providecommand \EOS [0]{\spacefactor3000\relax}%
\providecommand \BibitemShut  [1]{\csname bibitem#1\endcsname}%
\let\auto@bib@innerbib\@empty
\bibitem [{\citenamefont {Aaronson}(2013)}]{Aaronson}%
  \BibitemOpen
  \bibfield  {author} {\bibinfo {author} {\bibfnamefont {S.}~\bibnamefont
  {Aaronson}},\ }\href {https://doi.org/10.1017/CBO9780511979309} {\emph
  {\bibinfo {title} {Quantum Computing since Democritus}}}\ (\bibinfo
  {publisher} {Cambridge University Press},\ \bibinfo {year}
  {2013})\BibitemShut {NoStop}%
\bibitem [{\citenamefont {Akama}(2014)}]{Akama}%
  \BibitemOpen
  \bibfield  {author} {\bibinfo {author} {\bibfnamefont {S.}~\bibnamefont
  {Akama}},\ }\href {https://doi.org/https://doi.org/10.1007/978-3-319-08284-4}
  {\emph {\bibinfo {title} {Elements of Quantum Computing}}}\ (\bibinfo
  {publisher} {Springer Cham},\ \bibinfo {year} {2014})\BibitemShut {NoStop}%
\bibitem [{\citenamefont {Fischer}\ and\ \citenamefont
  {Loss}(2009)}]{Fischer2}%
  \BibitemOpen
  \bibfield  {author} {\bibinfo {author} {\bibfnamefont {J.}~\bibnamefont
  {Fischer}}\ and\ \bibinfo {author} {\bibfnamefont {D.}~\bibnamefont {Loss}},\
  }\bibfield  {title} {\bibinfo {title} {Dealing with decoherence},\ }\href
  {https://doi.org/10.1126/science.1169554} {\bibfield  {journal} {\bibinfo
  {journal} {Science}\ }\textbf {\bibinfo {volume} {324}},\ \bibinfo {pages}
  {1277} (\bibinfo {year} {2009})} \BibitemShut
  {NoStop}%
\bibitem [{\citenamefont {Zeh}(1970)}]{Zeh}%
  \BibitemOpen
  \bibfield  {author} {\bibinfo {author} {\bibfnamefont {H.~D.}\ \bibnamefont
  {Zeh}},\ }\bibfield  {title} {\bibinfo {title} {On the interpretation of
  measurement in quantum theory},\ }\href
  {https://link.springer.com/article/10.1007/BF00708656} {\bibfield  {journal}
  {\bibinfo  {journal} {Foundations of Physics}\ }\textbf {\bibinfo {volume}
  {1}},\ \bibinfo {pages} {69} (\bibinfo {year} {1970})}\BibitemShut {NoStop}%
\bibitem [{\citenamefont {Schlosshauer}(2019)}]{Schlosshauer}%
  \BibitemOpen
  \bibfield  {author} {\bibinfo {author} {\bibfnamefont {M.}~\bibnamefont
  {Schlosshauer}},\ }\bibfield  {title} {\bibinfo {title} {Quantum
  decoherence},\ }\href
  {https://doi.org/https://doi.org/10.1016/j.physrep.2019.10.001} {\bibfield
  {journal} {\bibinfo  {journal} {Physics Reports}\ }\textbf {\bibinfo {volume}
  {831}},\ \bibinfo {pages} {1} (\bibinfo {year} {2019})},\ \bibinfo {note}
  {quantum decoherence}\BibitemShut {NoStop}%
\bibitem [{\citenamefont {Zurek}(1981)}]{Zurek1}%
  \BibitemOpen
  \bibfield  {author} {\bibinfo {author} {\bibfnamefont {W.~H.}\ \bibnamefont
  {Zurek}},\ }\bibfield  {title} {\bibinfo {title} {Pointer basis of quantum
  apparatus: Into what mixture does the wave packet collapse?},\ }\href
  {https://doi.org/10.1103/PhysRevD.24.1516} {\bibfield  {journal} {\bibinfo
  {journal} {Phys. Rev. D}\ }\textbf {\bibinfo {volume} {24}},\ \bibinfo
  {pages} {1516} (\bibinfo {year} {1981})}\BibitemShut {NoStop}%
\bibitem [{\citenamefont {Zurek}(1982)}]{Zurek2}%
  \BibitemOpen
  \bibfield  {author} {\bibinfo {author} {\bibfnamefont {W.~H.}\ \bibnamefont
  {Zurek}},\ }\bibfield  {title} {\bibinfo {title} {Environment-induced
  superselection rules},\ }\href {https://doi.org/10.1103/PhysRevD.26.1862}
  {\bibfield  {journal} {\bibinfo  {journal} {Phys. Rev. D}\ }\textbf {\bibinfo
  {volume} {26}},\ \bibinfo {pages} {1862} (\bibinfo {year}
  {1982})}\BibitemShut {NoStop}%
\bibitem [{\citenamefont {Caldeira}\ and\ \citenamefont
  {Leggett}(1983{\natexlab{a}})}]{Caldeira1}%
  \BibitemOpen
  \bibfield  {author} {\bibinfo {author} {\bibfnamefont {A.}~\bibnamefont
  {Caldeira}}\ and\ \bibinfo {author} {\bibfnamefont {A.}~\bibnamefont
  {Leggett}},\ }\bibfield  {title} {\bibinfo {title} {Path integral approach to
  quantum brownian motion},\ }\href
  {https://doi.org/https://doi.org/10.1016/0378-4371(83)90013-4} {\bibfield
  {journal} {\bibinfo  {journal} {Physica A: Statistical Mechanics and its
  Applications}\ }\textbf {\bibinfo {volume} {121}},\ \bibinfo {pages} {587}
  (\bibinfo {year} {1983}{\natexlab{a}})}\BibitemShut {NoStop}%
\bibitem [{\citenamefont {Caldeira}\ and\ \citenamefont
  {Leggett}(1983{\natexlab{b}})}]{Caldeira2}%
  \BibitemOpen
  \bibfield  {author} {\bibinfo {author} {\bibfnamefont {A.}~\bibnamefont
  {Caldeira}}\ and\ \bibinfo {author} {\bibfnamefont {A.}~\bibnamefont
  {Leggett}},\ }\bibfield  {title} {\bibinfo {title} {Quantum tunnelling in a
  dissipative system},\ }\href
  {https://doi.org/https://doi.org/10.1016/0003-4916(83)90202-6} {\bibfield
  {journal} {\bibinfo  {journal} {Annals of Physics}\ }\textbf {\bibinfo
  {volume} {149}},\ \bibinfo {pages} {374} (\bibinfo {year}
  {1983}{\natexlab{b}})}\BibitemShut {NoStop}%
\bibitem [{\citenamefont {Unruh}\ and\ \citenamefont {Zurek}(1989)}]{Unruh}%
  \BibitemOpen
  \bibfield  {author} {\bibinfo {author} {\bibfnamefont {W.~G.}\ \bibnamefont
  {Unruh}}\ and\ \bibinfo {author} {\bibfnamefont {W.~H.}\ \bibnamefont
  {Zurek}},\ }\bibfield  {title} {\bibinfo {title} {Reduction of a wave packet
  in quantum brownian motion},\ }\href
  {https://doi.org/10.1103/PhysRevD.40.1071} {\bibfield  {journal} {\bibinfo
  {journal} {Phys. Rev. D}\ }\textbf {\bibinfo {volume} {40}},\ \bibinfo
  {pages} {1071} (\bibinfo {year} {1989})}\BibitemShut {NoStop}%
\bibitem [{\citenamefont {Hu}\ \emph {et~al.}(1992)\citenamefont {Hu},
  \citenamefont {Paz},\ and\ \citenamefont {Zhang}}]{HuPaz}%
  \BibitemOpen
  \bibfield  {author} {\bibinfo {author} {\bibfnamefont {B.~L.}\ \bibnamefont
  {Hu}}, \bibinfo {author} {\bibfnamefont {J.~P.}\ \bibnamefont {Paz}},\ and\
  \bibinfo {author} {\bibfnamefont {Y.}~\bibnamefont {Zhang}},\ }\bibfield
  {title} {\bibinfo {title} {Quantum brownian motion in a general environment:
  Exact master equation with nonlocal dissipation and colored noise},\ }\href
  {https://doi.org/10.1103/PhysRevD.45.2843} {\bibfield  {journal} {\bibinfo
  {journal} {Phys. Rev. D}\ }\textbf {\bibinfo {volume} {45}},\ \bibinfo
  {pages} {2843} (\bibinfo {year} {1992})}\BibitemShut {NoStop}%
\bibitem [{\citenamefont {Brun}\ \emph {et~al.}(2003)\citenamefont {Brun},
  \citenamefont {Carteret},\ and\ \citenamefont {Ambainis}}]{Brun}%
  \BibitemOpen
  \bibfield  {author} {\bibinfo {author} {\bibfnamefont {T.~A.}\ \bibnamefont
  {Brun}}, \bibinfo {author} {\bibfnamefont {H.~A.}\ \bibnamefont {Carteret}},\
  and\ \bibinfo {author} {\bibfnamefont {A.}~\bibnamefont {Ambainis}},\
  }\bibfield  {title} {\bibinfo {title} {Quantum to classical transition for
  random walks},\ }\href {https://doi.org/10.1103/PhysRevLett.91.130602}
  {\bibfield  {journal} {\bibinfo  {journal} {Phys. Rev. Lett.}\ }\textbf
  {\bibinfo {volume} {91}},\ \bibinfo {pages} {130602} (\bibinfo {year}
  {2003})}\BibitemShut {NoStop}%
\bibitem [{\citenamefont {H.N.Xiong}\ \emph {et~al.}(2015)\citenamefont
  {H.N.Xiong}, \citenamefont {P.Y.Lo}, \citenamefont {W.M.Zhang}, \citenamefont
  {D.H.Feng},\ and\ \citenamefont {F.Nori}}]{Xiong}%
  \BibitemOpen
  \bibfield  {author} {\bibinfo {author} {\bibnamefont {H.N.Xiong}}, \bibinfo
  {author} {\bibnamefont {P.Y.Lo}}, \bibinfo {author} {\bibnamefont
  {W.M.Zhang}}, \bibinfo {author} {\bibnamefont {D.H.Feng}},\ and\ \bibinfo
  {author} {\bibnamefont {F.Nori}},\ }\bibfield  {title} {\bibinfo {title}
  {Non-markovian complexity in the quantum-to-classical transition},\ }\href
  {https://doi.org/10.1038/srep13353} {\bibfield  {journal} {\bibinfo
  {journal} {Sci. Rep.}\ }\textbf {\bibinfo {volume} {13353}} (\bibinfo {year}
  {2015})}\BibitemShut {NoStop}%
\bibitem [{\citenamefont {Zurek}(2007)}]{Zurek3}%
  \BibitemOpen
  \bibfield  {author} {\bibinfo {author} {\bibfnamefont {W.~H.}\ \bibnamefont
  {Zurek}},\ }\bibinfo {title} {Decoherence and the transition from quantum to
  classical --- revisited},\ in\ \href
  {https://doi.org/10.1007/978-3-7643-7808-0_1} {\emph {\bibinfo {booktitle}
  {Quantum Decoherence: Poincar{\'e} Seminar 2005}}},\ \bibinfo {editor}
  {edited by\ \bibinfo {editor} {\bibfnamefont {B.}~\bibnamefont {Duplantier}},
  \bibinfo {editor} {\bibfnamefont {J.-M.}\ \bibnamefont {Raimond}},\ and\
  \bibinfo {editor} {\bibfnamefont {V.}~\bibnamefont {Rivasseau}}}\ (\bibinfo
  {publisher} {Birkh{\"a}user Basel},\ \bibinfo {address} {Basel},\ \bibinfo
  {year} {2007})\ pp.\ \bibinfo {pages} {1--31}\BibitemShut {NoStop}%
\bibitem [{\citenamefont {Orszag}(2016)}]{Orszag}%
  \BibitemOpen
  \bibfield  {author} {\bibinfo {author} {\bibfnamefont {M.}~\bibnamefont
  {Orszag}},\ }\href@noop {} {\emph {\bibinfo {title} {Quantum optics:
  including noise reduction, trapped ions, quantum trajectories, and
  decoherence}}}\ (\bibinfo  {publisher} {Springer},\ \bibinfo {year}
  {2016})\BibitemShut {NoStop}%
\bibitem [{\citenamefont {Cucchietti}\ \emph {et~al.}(2005)\citenamefont
  {Cucchietti}, \citenamefont {Paz},\ and\ \citenamefont {Zurek}}]{Cucchietti}%
  \BibitemOpen
  \bibfield  {author} {\bibinfo {author} {\bibfnamefont {F.~M.}\ \bibnamefont
  {Cucchietti}}, \bibinfo {author} {\bibfnamefont {J.~P.}\ \bibnamefont
  {Paz}},\ and\ \bibinfo {author} {\bibfnamefont {W.~H.}\ \bibnamefont
  {Zurek}},\ }\bibfield  {title} {\bibinfo {title} {Decoherence from spin
  environments},\ }\href {https://doi.org/10.1103/PhysRevA.72.052113}
  {\bibfield  {journal} {\bibinfo  {journal} {Phys. Rev. A}\ }\textbf {\bibinfo
  {volume} {72}},\ \bibinfo {pages} {052113} (\bibinfo {year}
  {2005})}\BibitemShut {NoStop}%
\bibitem [{\citenamefont {Devoret}\ and\ \citenamefont
  {Schoelkopf}(2013)}]{Devoret}%
  \BibitemOpen
  \bibfield  {author} {\bibinfo {author} {\bibfnamefont {M.~H.}\ \bibnamefont
  {Devoret}}\ and\ \bibinfo {author} {\bibfnamefont {R.~J.}\ \bibnamefont
  {Schoelkopf}},\ }\bibfield  {title} {\bibinfo {title} {Superconducting
  circuits for quantum information: An outlook},\ }\href
  {https://doi.org/10.1126/science.1231930} {\bibfield  {journal} {\bibinfo
  {journal} {Science}\ }\textbf {\bibinfo {volume} {339}},\ \bibinfo {pages}
  {1169} (\bibinfo {year} {2013})},\ \Eprint
  {https://arxiv.org/abs/https://www.science.org/doi/pdf/10.1126/science.1231930}
  {https://www.science.org/doi/pdf/10.1126/science.1231930} \BibitemShut
  {NoStop}%
\bibitem [{\citenamefont {Makhlin}\ \emph {et~al.}(2001)\citenamefont
  {Makhlin}, \citenamefont {Sch\"on},\ and\ \citenamefont
  {Shnirman}}]{Makhlin}%
  \BibitemOpen
  \bibfield  {author} {\bibinfo {author} {\bibfnamefont {Y.}~\bibnamefont
  {Makhlin}}, \bibinfo {author} {\bibfnamefont {G.}~\bibnamefont {Sch\"on}},\
  and\ \bibinfo {author} {\bibfnamefont {A.}~\bibnamefont {Shnirman}},\
  }\bibfield  {title} {\bibinfo {title} {Quantum-state engineering with
  josephson-junction devices},\ }\href
  {https://doi.org/10.1103/RevModPhys.73.357} {\bibfield  {journal} {\bibinfo
  {journal} {Rev. Mod. Phys.}\ }\textbf {\bibinfo {volume} {73}},\ \bibinfo
  {pages} {357} (\bibinfo {year} {2001})}\BibitemShut {NoStop}%
\bibitem [{\citenamefont {Sinha}(1997)}]{Sinhadecoherence}%
  \BibitemOpen
  \bibfield  {author} {\bibinfo {author} {\bibfnamefont {S.}~\bibnamefont
  {Sinha}},\ }\bibfield  {title} {\bibinfo {title} {Decoherence at absolute
  zero},\ }\href
  {https://doi.org/https://doi.org/10.1016/S0375-9601(97)00098-4} {\bibfield
  {journal} {\bibinfo  {journal} {Physics Letters A}\ }\textbf {\bibinfo
  {volume} {228}},\ \bibinfo {pages} {1} (\bibinfo {year} {1997})}\BibitemShut
  {NoStop}%
\bibitem [{\citenamefont {Ford}\ and\ \citenamefont
  {O’Connell}(2003)}]{Ford}%
  \BibitemOpen
  \bibfield  {author} {\bibinfo {author} {\bibfnamefont {G.~W.}\ \bibnamefont
  {Ford}}\ and\ \bibinfo {author} {\bibfnamefont {R.~F.}\ \bibnamefont
  {O’Connell}},\ }\bibfield  {title} {\bibinfo {title} {Decoherence at zero
  temperature},\ }\href {https://doi.org/10.1088/1464-4266/5/6/010} {\bibfield
  {journal} {\bibinfo  {journal} {Journal of Optics B: Quantum and
  Semiclassical Optics}\ }\textbf {\bibinfo {volume} {5}},\ \bibinfo {pages}
  {S609} (\bibinfo {year} {2003})}\BibitemShut {NoStop}%
\bibitem [{\citenamefont {Mohanty}(2000)}]{Mohanty}%
  \BibitemOpen
  \bibfield  {author} {\bibinfo {author} {\bibfnamefont {P.}~\bibnamefont
  {Mohanty}},\ }\bibfield  {title} {\bibinfo {title} {Notes on decoherence at
  absolute zero},\ }\href
  {https://doi.org/https://doi.org/10.1016/S0921-4526(99)01833-5} {\bibfield
  {journal} {\bibinfo  {journal} {Physica B: Condensed Matter}\ }\textbf
  {\bibinfo {volume} {280}},\ \bibinfo {pages} {446} (\bibinfo {year}
  {2000})}\BibitemShut {NoStop}%
\bibitem [{\citenamefont {Lombardo}\ and\ \citenamefont
  {Villar}(2005)}]{Lombardo}%
  \BibitemOpen
  \bibfield  {author} {\bibinfo {author} {\bibfnamefont {F.~C.}\ \bibnamefont
  {Lombardo}}\ and\ \bibinfo {author} {\bibfnamefont {P.~I.}\ \bibnamefont
  {Villar}},\ }\bibfield  {title} {\bibinfo {title} {Decoherence induced by
  zero-point fluctuations in quantum brownian motion},\ }\href
  {https://doi.org/https://doi.org/10.1016/j.physleta.2004.12.065} {\bibfield
  {journal} {\bibinfo  {journal} {Physics Letters A}\ }\textbf {\bibinfo
  {volume} {336}},\ \bibinfo {pages} {16} (\bibinfo {year} {2005})}\BibitemShut
  {NoStop}%
\bibitem [{\citenamefont {Mohanty}\ and\ \citenamefont
  {Webb}(1997)}]{Mohanty2}%
  \BibitemOpen
  \bibfield  {author} {\bibinfo {author} {\bibfnamefont {P.}~\bibnamefont
  {Mohanty}}\ and\ \bibinfo {author} {\bibfnamefont {R.~A.}\ \bibnamefont
  {Webb}},\ }\bibfield  {title} {\bibinfo {title} {Decoherence and quantum
  fluctuations},\ }\href {https://doi.org/10.1103/PhysRevB.55.R13452}
  {\bibfield  {journal} {\bibinfo  {journal} {Phys. Rev. B}\ }\textbf {\bibinfo
  {volume} {55}},\ \bibinfo {pages} {R13452} (\bibinfo {year}
  {1997})}\BibitemShut {NoStop}%
\bibitem [{\citenamefont {Ren}\ \emph {et~al.}(2013)\citenamefont {Ren},
  \citenamefont {Heremans}, \citenamefont {Gaspe}, \citenamefont
  {Vijeyaragunathan}, \citenamefont {Mishima},\ and\ \citenamefont
  {Santos}}]{Ren}%
  \BibitemOpen
  \bibfield  {author} {\bibinfo {author} {\bibfnamefont {S.~L.}\ \bibnamefont
  {Ren}}, \bibinfo {author} {\bibfnamefont {J.~J.}\ \bibnamefont {Heremans}},
  \bibinfo {author} {\bibfnamefont {C.~K.}\ \bibnamefont {Gaspe}}, \bibinfo
  {author} {\bibfnamefont {S.}~\bibnamefont {Vijeyaragunathan}}, \bibinfo
  {author} {\bibfnamefont {T.~D.}\ \bibnamefont {Mishima}},\ and\ \bibinfo
  {author} {\bibfnamefont {M.~B.}\ \bibnamefont {Santos}},\ }\bibfield  {title}
  {\bibinfo {title} {Aharonov–bohm oscillations, quantum decoherence and
  amplitude modulation in mesoscopic ingaas/inalas rings},\ }\href
  {https://doi.org/10.1088/0953-8984/25/43/435301} {\bibfield  {journal}
  {\bibinfo  {journal} {Journal of Physics: Condensed Matter}\ }\textbf
  {\bibinfo {volume} {25}},\ \bibinfo {pages} {435301} (\bibinfo {year}
  {2013})}\BibitemShut {NoStop}%
\bibitem [{\citenamefont {Giulio}\ and\ \citenamefont
  {de~Abajo}(2020)}]{Giulio}%
  \BibitemOpen
  \bibfield  {author} {\bibinfo {author} {\bibfnamefont {V.~D.}\ \bibnamefont
  {Giulio}}\ and\ \bibinfo {author} {\bibfnamefont {F.~J.~G.}\ \bibnamefont
  {de~Abajo}},\ }\bibfield  {title} {\bibinfo {title} {Electron diffraction by
  vacuum fluctuations},\ }\href {https://doi.org/10.1088/1367-2630/abbddf}
  {\bibfield  {journal} {\bibinfo  {journal} {New Journal of Physics}\ }\textbf
  {\bibinfo {volume} {22}},\ \bibinfo {pages} {103057} (\bibinfo {year}
  {2020})}\BibitemShut {NoStop}%
\bibitem [{\citenamefont {Hanson}\ \emph {et~al.}(2007)\citenamefont {Hanson},
  \citenamefont {Kouwenhoven}, \citenamefont {Petta}, \citenamefont {Tarucha},\
  and\ \citenamefont {Vandersypen}}]{Hanson}%
  \BibitemOpen
  \bibfield  {author} {\bibinfo {author} {\bibfnamefont {R.}~\bibnamefont
  {Hanson}}, \bibinfo {author} {\bibfnamefont {L.~P.}\ \bibnamefont
  {Kouwenhoven}}, \bibinfo {author} {\bibfnamefont {J.~R.}\ \bibnamefont
  {Petta}}, \bibinfo {author} {\bibfnamefont {S.}~\bibnamefont {Tarucha}},\
  and\ \bibinfo {author} {\bibfnamefont {L.~M.~K.}\ \bibnamefont
  {Vandersypen}},\ }\bibfield  {title} {\bibinfo {title} {Spins in few-electron
  quantum dots},\ }\href {https://doi.org/10.1103/RevModPhys.79.1217}
  {\bibfield  {journal} {\bibinfo  {journal} {Rev. Mod. Phys.}\ }\textbf
  {\bibinfo {volume} {79}},\ \bibinfo {pages} {1217} (\bibinfo {year}
  {2007})}\BibitemShut {NoStop}%
\bibitem [{\citenamefont {Fischer}\ \emph {et~al.}(2008)\citenamefont
  {Fischer}, \citenamefont {Coish}, \citenamefont {Bulaev},\ and\ \citenamefont
  {Loss}}]{Fischer}%
  \BibitemOpen
  \bibfield  {author} {\bibinfo {author} {\bibfnamefont {J.}~\bibnamefont
  {Fischer}}, \bibinfo {author} {\bibfnamefont {W.~A.}\ \bibnamefont {Coish}},
  \bibinfo {author} {\bibfnamefont {D.~V.}\ \bibnamefont {Bulaev}},\ and\
  \bibinfo {author} {\bibfnamefont {D.}~\bibnamefont {Loss}},\ }\bibfield
  {title} {\bibinfo {title} {Spin decoherence of a heavy hole coupled to
  nuclear spins in a quantum dot},\ }\href
  {https://doi.org/10.1103/PhysRevB.78.155329} {\bibfield  {journal} {\bibinfo
  {journal} {Phys. Rev. B}\ }\textbf {\bibinfo {volume} {78}},\ \bibinfo
  {pages} {155329} (\bibinfo {year} {2008})}\BibitemShut {NoStop}%
\bibitem [{\citenamefont {Hanson}\ and\ \citenamefont
  {Awschalom}(2008)}]{Hanson2}%
  \BibitemOpen
  \bibfield  {author} {\bibinfo {author} {\bibfnamefont {R.}~\bibnamefont
  {Hanson}}\ and\ \bibinfo {author} {\bibfnamefont {D.~D.}\ \bibnamefont
  {Awschalom}},\ }\bibfield  {title} {\bibinfo {title} {Coherent manipulation
  of single spins in semiconductors},\ }\href
  {https://www.nature.com/articles/nature07129} {\bibfield  {journal} {\bibinfo
   {journal} {Nature}\ }\textbf {\bibinfo {volume} {453}},\ \bibinfo {pages}
  {1043} (\bibinfo {year} {2008})}\BibitemShut {NoStop}%
\bibitem [{\citenamefont {Beau}\ \emph {et~al.}(2017)\citenamefont {Beau},
  \citenamefont {Kiukas}, \citenamefont {Egusquiza},\ and\ \citenamefont {del
  Campo}}]{Beau}%
  \BibitemOpen
  \bibfield  {author} {\bibinfo {author} {\bibfnamefont {M.}~\bibnamefont
  {Beau}}, \bibinfo {author} {\bibfnamefont {J.}~\bibnamefont {Kiukas}},
  \bibinfo {author} {\bibfnamefont {I.~L.}\ \bibnamefont {Egusquiza}},\ and\
  \bibinfo {author} {\bibfnamefont {A.}~\bibnamefont {del Campo}},\ }\bibfield
  {title} {\bibinfo {title} {Nonexponential quantum decay under environmental
  decoherence},\ }\href {https://doi.org/10.1103/PhysRevLett.119.130401}
  {\bibfield  {journal} {\bibinfo  {journal} {Phys. Rev. Lett.}\ }\textbf
  {\bibinfo {volume} {119}},\ \bibinfo {pages} {130401} (\bibinfo {year}
  {2017})}\BibitemShut {NoStop}%
\bibitem [{\citenamefont {Polonyi}(2018)}]{Polonyi}%
  \BibitemOpen
  \bibfield  {author} {\bibinfo {author} {\bibfnamefont {J.}~\bibnamefont
  {Polonyi}},\ }\bibfield  {title} {\bibinfo {title} {Instantaneous and
  dynamical decoherence},\ }\href {https://doi.org/10.1088/1751-8121/aab0ad}
  {\bibfield  {journal} {\bibinfo  {journal} {Journal of Physics A:
  Mathematical and Theoretical}\ }\textbf {\bibinfo {volume} {51}},\ \bibinfo
  {pages} {145302} (\bibinfo {year} {2018})}\BibitemShut {NoStop}%
\bibitem [{\citenamefont {Schlosshauer}\ \emph {et~al.}(2008)\citenamefont
  {Schlosshauer}, \citenamefont {Hines},\ and\ \citenamefont
  {Milburn}}]{Schlosshauertwolevel}%
  \BibitemOpen
  \bibfield  {author} {\bibinfo {author} {\bibfnamefont {M.}~\bibnamefont
  {Schlosshauer}}, \bibinfo {author} {\bibfnamefont {A.~P.}\ \bibnamefont
  {Hines}},\ and\ \bibinfo {author} {\bibfnamefont {G.~J.}\ \bibnamefont
  {Milburn}},\ }\bibfield  {title} {\bibinfo {title} {Decoherence and
  dissipation of a quantum harmonic oscillator coupled to two-level systems},\
  }\href {https://doi.org/10.1103/PhysRevA.77.022111} {\bibfield  {journal}
  {\bibinfo  {journal} {Phys. Rev. A}\ }\textbf {\bibinfo {volume} {77}},\
  \bibinfo {pages} {022111} (\bibinfo {year} {2008})}\BibitemShut {NoStop}%
\bibitem [{\citenamefont {Carlesso}\ and\ \citenamefont
  {Bassi}(2017)}]{Carlesso}%
  \BibitemOpen
  \bibfield  {author} {\bibinfo {author} {\bibfnamefont {M.}~\bibnamefont
  {Carlesso}}\ and\ \bibinfo {author} {\bibfnamefont {A.}~\bibnamefont
  {Bassi}},\ }\bibfield  {title} {\bibinfo {title} {Adjoint master equation for
  quantum brownian motion},\ }\href
  {https://doi.org/10.1103/PhysRevA.95.052119} {\bibfield  {journal} {\bibinfo
  {journal} {Phys. Rev. A}\ }\textbf {\bibinfo {volume} {95}},\ \bibinfo
  {pages} {052119} (\bibinfo {year} {2017})}\BibitemShut {NoStop}%
\bibitem [{\citenamefont {Hörhammer}\ and\ \citenamefont
  {Büttner}(2008)}]{Horhammer}%
  \BibitemOpen
  \bibfield  {author} {\bibinfo {author} {\bibfnamefont {C.}~\bibnamefont
  {Hörhammer}}\ and\ \bibinfo {author} {\bibfnamefont {H.}~\bibnamefont
  {Büttner}},\ }\bibfield  {title} {\bibinfo {title} {Decoherence and
  disentanglement scenarios in non-markovian quantum brownian motion},\ }\href
  {https://doi.org/10.1088/1751-8113/41/26/265301} {\bibfield  {journal}
  {\bibinfo  {journal} {Journal of Physics A: Mathematical and Theoretical}\
  }\textbf {\bibinfo {volume} {41}},\ \bibinfo {pages} {265301} (\bibinfo
  {year} {2008})}\BibitemShut {NoStop}%
\bibitem [{\citenamefont {Ghorashi}\ and\ \citenamefont
  {Harouni}(2013)}]{Ghorashi}%
  \BibitemOpen
  \bibfield  {author} {\bibinfo {author} {\bibfnamefont {S.}~\bibnamefont
  {Ghorashi}}\ and\ \bibinfo {author} {\bibfnamefont {M.~B.}\ \bibnamefont
  {Harouni}},\ }\bibfield  {title} {\bibinfo {title} {Decoherence of quantum
  brownian motion in noncommutative space},\ }\href
  {https://doi.org/https://doi.org/10.1016/j.physleta.2013.02.019} {\bibfield
  {journal} {\bibinfo  {journal} {Physics Letters A}\ }\textbf {\bibinfo
  {volume} {377}},\ \bibinfo {pages} {952} (\bibinfo {year}
  {2013})}\BibitemShut {NoStop}%
\bibitem [{\citenamefont {Tcoffo}\ \emph {et~al.}(2020)\citenamefont {Tcoffo},
  \citenamefont {Deuto}, \citenamefont {Nsangou}, \citenamefont {Koumetio},
  \citenamefont {Tchapda},\ and\ \citenamefont {Tene}}]{Tcoffo}%
  \BibitemOpen
  \bibfield  {author} {\bibinfo {author} {\bibfnamefont {M.}~\bibnamefont
  {Tcoffo}}, \bibinfo {author} {\bibfnamefont {G.}~\bibnamefont {Deuto}},
  \bibinfo {author} {\bibfnamefont {I.}~\bibnamefont {Nsangou}}, \bibinfo
  {author} {\bibfnamefont {A.}~\bibnamefont {Koumetio}}, \bibinfo {author}
  {\bibfnamefont {L.}~\bibnamefont {Tchapda}},\ and\ \bibinfo {author}
  {\bibfnamefont {A.}~\bibnamefont {Tene}},\ }\bibfield  {title} {\bibinfo
  {title} {Decoherence of a damped anisotropic harmonic oscillator under
  magnetic field effects in a two-dimensional noncommutative phase-space},\
  }\href {https://doi.org/10.4236/jamp.2020.812207} {\bibfield  {journal}
  {\bibinfo  {journal} {Journal of Applied Mathematics and Physics}\ }\textbf
  {\bibinfo {volume} {8}},\ \bibinfo {pages} {2801} (\bibinfo {year}
  {2020})}\BibitemShut {NoStop}%
\bibitem [{\citenamefont {Germain}\ \emph {et~al.}(2021)\citenamefont
  {Germain}, \citenamefont {Armel}, \citenamefont {Tene}, \citenamefont
  {Isofa},\ and\ \citenamefont {Tchoffo}}]{Germain}%
  \BibitemOpen
  \bibfield  {author} {\bibinfo {author} {\bibfnamefont {Y.~D.}\ \bibnamefont
  {Germain}}, \bibinfo {author} {\bibfnamefont {A.~K.}\ \bibnamefont {Armel}},
  \bibinfo {author} {\bibfnamefont {A.~G.}\ \bibnamefont {Tene}}, \bibinfo
  {author} {\bibfnamefont {N.}~\bibnamefont {Isofa}},\ and\ \bibinfo {author}
  {\bibfnamefont {M.}~\bibnamefont {Tchoffo}},\ }\bibfield  {title} {\bibinfo
  {title} {Decoherence dynamics of a charged particle within a non-demolition
  type interaction in non-commutative phase-space},\ }\href
  {https://doi.org/10.1088/1402-4896/ac0273} {\bibfield  {journal} {\bibinfo
  {journal} {Physica Scripta}\ }\textbf {\bibinfo {volume} {96}},\ \bibinfo
  {pages} {085705} (\bibinfo {year} {2021})}\BibitemShut {NoStop}%
\bibitem [{\citenamefont {Armel}\ \emph {et~al.}(2021)\citenamefont {Armel},
  \citenamefont {Germain}, \citenamefont {Giresse},\ and\ \citenamefont
  {Martin}}]{Armel}%
  \BibitemOpen
  \bibfield  {author} {\bibinfo {author} {\bibfnamefont {A.~K.}\ \bibnamefont
  {Armel}}, \bibinfo {author} {\bibfnamefont {Y.~D.}\ \bibnamefont {Germain}},
  \bibinfo {author} {\bibfnamefont {T.~A.}\ \bibnamefont {Giresse}},\ and\
  \bibinfo {author} {\bibfnamefont {T.}~\bibnamefont {Martin}},\ }\bibfield
  {title} {\bibinfo {title} {The dynamic of quantum entanglement of two
  dimensional harmonic oscillator in non-commutative space},\ }\href
  {https://doi.org/10.1088/1402-4896/ac42a9} {\bibfield  {journal} {\bibinfo
  {journal} {Physica Scripta}\ }\textbf {\bibinfo {volume} {96}},\ \bibinfo
  {pages} {125731} (\bibinfo {year} {2021})}\BibitemShut {NoStop}%
\bibitem [{\citenamefont {Ferialdi}\ and\ \citenamefont
  {Smirne}(2017)}]{Ferialdi}%
  \BibitemOpen
  \bibfield  {author} {\bibinfo {author} {\bibfnamefont {L.}~\bibnamefont
  {Ferialdi}}\ and\ \bibinfo {author} {\bibfnamefont {A.}~\bibnamefont
  {Smirne}},\ }\bibfield  {title} {\bibinfo {title} {Momentum coupling in
  non-markovian quantum brownian motion},\ }\href
  {https://doi.org/10.1103/PhysRevA.96.012109} {\bibfield  {journal} {\bibinfo
  {journal} {Phys. Rev. A}\ }\textbf {\bibinfo {volume} {96}},\ \bibinfo
  {pages} {012109} (\bibinfo {year} {2017})}\BibitemShut {NoStop}%
\bibitem [{\citenamefont {Schlosshauer}(2007)}]{Maximilianbook}%
  \BibitemOpen
  \bibfield  {author} {\bibinfo {author} {\bibfnamefont {M.}~\bibnamefont
  {Schlosshauer}},\ }\href {https://doi.org/10.1007/978-3-540-35775-9} {\emph
  {\bibinfo {title} {Decoherence and the Quantum-To-Classical Transition}}},\
  edited by\ \bibinfo {editor} {\bibnamefont {A.C.Elitzur}}, \bibinfo {editor}
  {\bibnamefont {M.P.Silverman}}, \bibinfo {editor} {\bibnamefont
  {J.Tuszynski}}, \bibinfo {editor} {\bibnamefont {R.Vaas}},\ and\ \bibinfo
  {editor} {\bibnamefont {H.D.Zeh}}\ (\bibinfo  {publisher} {Springer Berlin,
  Heidelberg},\ \bibinfo {year} {2007})\BibitemShut {NoStop}%
\bibitem [{\citenamefont {Breuer}\ \emph {et~al.}(2002)\citenamefont {Breuer},
  \citenamefont {Petruccione} \emph {et~al.}}]{Breuer}%
  \BibitemOpen
  \bibfield  {author} {\bibinfo {author} {\bibfnamefont {H.-P.}\ \bibnamefont
  {Breuer}}, \bibinfo {author} {\bibfnamefont {F.}~\bibnamefont {Petruccione}},
  \emph {et~al.},\ }\href@noop {} {\emph {\bibinfo {title} {The theory of open
  quantum systems}}}\ (\bibinfo  {publisher} {Oxford University Press on
  Demand},\ \bibinfo {year} {2002})\BibitemShut {NoStop}%
\bibitem [{\citenamefont {Redfield}(1957)}]{Redfield}%
  \BibitemOpen
  \bibfield  {author} {\bibinfo {author} {\bibfnamefont {A.~G.}\ \bibnamefont
  {Redfield}},\ }\bibfield  {title} {\bibinfo {title} {On the theory of
  relaxation processes},\ }\href {https://doi.org/10.1147/rd.11.0019}
  {\bibfield  {journal} {\bibinfo  {journal} {IBM Journal of Research and
  Development}\ }\textbf {\bibinfo {volume} {1}},\ \bibinfo {pages} {19}
  (\bibinfo {year} {1957})}\BibitemShut {NoStop}%
\bibitem [{\citenamefont {Blum}(2012)}]{Blum}%
  \BibitemOpen
  \bibfield  {author} {\bibinfo {author} {\bibfnamefont {K.}~\bibnamefont
  {Blum}},\ }\href
  {https://books.google.co.in/books?hl=en&lr=&id=o0Bofi3_ZI0C&oi=fnd&pg=PR3&dq=Blum+2012&ots=pkzZ-E4WO1&sig=WRtzASv7gVEm5OaiTdh8HHhFZbo&redir_esc=y#v=onepage&q=Blum%202012&f=false}
  {\emph {\bibinfo {title} {Density matrix theory and applications}}},\
  Vol.~\bibinfo {volume} {64}\ (\bibinfo  {publisher} {Springer Science \&
  Business Media},\ \bibinfo {year} {2012})\BibitemShut {NoStop}%
\bibitem [{\citenamefont {Caldeira}\ and\ \citenamefont
  {Leggett}(1985)}]{Caldeira3}%
  \BibitemOpen
  \bibfield  {author} {\bibinfo {author} {\bibfnamefont {A.}~\bibnamefont
  {Caldeira}}\ and\ \bibinfo {author} {\bibfnamefont {A.~J.}\ \bibnamefont
  {Leggett}},\ }\bibfield  {title} {\bibinfo {title} {Influence of damping on
  quantum interference: An exactly soluble model},\ }\href@noop {} {\bibfield
  {journal} {\bibinfo  {journal} {Physical Review A}\ }\textbf {\bibinfo
  {volume} {31}},\ \bibinfo {pages} {1059} (\bibinfo {year}
  {1985})}\BibitemShut {NoStop}%
\bibitem [{\citenamefont {Intravaia}\ \emph {et~al.}(2003)\citenamefont
  {Intravaia}, \citenamefont {Maniscalco},\ and\ \citenamefont
  {Messina}}]{Intravaia}%
  \BibitemOpen
  \bibfield  {author} {\bibinfo {author} {\bibfnamefont {F.}~\bibnamefont
  {Intravaia}}, \bibinfo {author} {\bibfnamefont {S.}~\bibnamefont
  {Maniscalco}},\ and\ \bibinfo {author} {\bibfnamefont {A.}~\bibnamefont
  {Messina}},\ }\bibfield  {title} {\bibinfo {title} {Density-matrix
  operatorial solution of the non-markovian master equation for quantum
  brownian motion},\ }\href
  {https://journals.aps.org/pra/abstract/10.1103/PhysRevA.67.042108} {\bibfield
   {journal} {\bibinfo  {journal} {Physical Review A}\ }\textbf {\bibinfo
  {volume} {67}},\ \bibinfo {pages} {042108} (\bibinfo {year}
  {2003})}\BibitemShut {NoStop}%
\bibitem [{\citenamefont {Sarkar}\ \emph {et~al.}(2017)\citenamefont {Sarkar},
  \citenamefont {Paul}, \citenamefont {Vishwakarma}, \citenamefont {Kumar},
  \citenamefont {Verma}, \citenamefont {Sainath}, \citenamefont {Rapol},\ and\
  \citenamefont {Santhanam}}]{Sarkar}%
  \BibitemOpen
  \bibfield  {author} {\bibinfo {author} {\bibfnamefont {S.}~\bibnamefont
  {Sarkar}}, \bibinfo {author} {\bibfnamefont {S.}~\bibnamefont {Paul}},
  \bibinfo {author} {\bibfnamefont {C.}~\bibnamefont {Vishwakarma}}, \bibinfo
  {author} {\bibfnamefont {S.}~\bibnamefont {Kumar}}, \bibinfo {author}
  {\bibfnamefont {G.}~\bibnamefont {Verma}}, \bibinfo {author} {\bibfnamefont
  {M.}~\bibnamefont {Sainath}}, \bibinfo {author} {\bibfnamefont {U.~D.}\
  \bibnamefont {Rapol}},\ and\ \bibinfo {author} {\bibfnamefont {M.~S.}\
  \bibnamefont {Santhanam}},\ }\bibfield  {title} {\bibinfo {title}
  {Nonexponential decoherence and subdiffusion in atom-optics kicked rotor},\
  }\href {https://doi.org/10.1103/PhysRevLett.118.174101} {\bibfield  {journal}
  {\bibinfo  {journal} {Phys. Rev. Lett.}\ }\textbf {\bibinfo {volume} {118}},\
  \bibinfo {pages} {174101} (\bibinfo {year} {2017})}\BibitemShut {NoStop}%
\bibitem [{\citenamefont {Lerch}(1900)}]{Lerch}%
  \BibitemOpen
  \bibfield  {author} {\bibinfo {author} {\bibfnamefont {M.}~\bibnamefont
  {Lerch}},\ }\bibfield  {title} {\bibinfo {title} {{Note sur la fonction
  $\mathfrak{K}(w, x, s) = \sum\limits_{k = 0}^\infty {\frac{{e^{2k\pi ix}
  }}{{\left( {w + k} \right)^s }}} $}},\ }\href
  {https://doi.org/10.1007/BF02612318} {\bibfield  {journal} {\bibinfo
  {journal} {Acta Mathematica}\ }\textbf {\bibinfo {volume} {11}},\ \bibinfo
  {pages} {19 } (\bibinfo {year} {1900})}\BibitemShut {NoStop}%
\bibitem [{\citenamefont {Andrews}(1998)}]{Andrews}%
  \BibitemOpen
  \bibfield  {author} {\bibinfo {author} {\bibfnamefont {L.~C.}\ \bibnamefont
  {Andrews}},\ }\href
  {https://books.google.co.in/books?hl=en&lr=&id=2CAqsF-RebgC&oi=fnd&pg=PR11&dq=Andrews,+Larry+C.+(1998).+Special+functions+of+mathematics+for+engineers.+SPIE+Press.+p.+110.+ISBN+9780819426161.&ots=LC4ZLSoI37&sig=Xtn8sCbtJrLh0CrfYF-kO8RdJG0&redir_esc=y#v=onepage&q=Andrews%2C%20Larry%20C.%20(1998).%20Special%20functions%20of%20mathematics%20for%20engineers.%20SPIE%20Press.%20p.%20110.%20ISBN%209780819426161.&f=false}
  {\emph {\bibinfo {title} {Special functions of mathematics for engineers}}},\
  Vol.~\bibinfo {volume} {49}\ (\bibinfo  {publisher} {Spie Press},\ \bibinfo
  {year} {1998})\BibitemShut {NoStop}%
\bibitem [{\citenamefont {Szeg{\"o}}(1954)}]{Szeg}%
  \BibitemOpen
  \bibfield  {author} {\bibinfo {author} {\bibfnamefont {G.}~\bibnamefont
  {Szeg{\"o}}},\ }\bibfield  {title} {\bibinfo {title} {A. erd{\'e}lyi, w.
  magnus, f. oberhettinger and fg tricomi, higher transcendental functions},\
  }\href
  {https://www.ams.org/journals/bull/1954-60-04/S0002-9904-1954-09835-X/S0002-9904-1954-09835-X.pdf}
  {\bibfield  {journal} {\bibinfo  {journal} {Bulletin of the American
  Mathematical Society}\ }\textbf {\bibinfo {volume} {60}},\ \bibinfo {pages}
  {405} (\bibinfo {year} {1954})}\BibitemShut {NoStop}%
\bibitem [{\citenamefont {Almog}\ \emph {et~al.}(2011)\citenamefont {Almog},
  \citenamefont {Sagi}, \citenamefont {Gordon}, \citenamefont {Bensky},
  \citenamefont {Kurizki},\ and\ \citenamefont {Davidson}}]{Almog}%
  \BibitemOpen
  \bibfield  {author} {\bibinfo {author} {\bibfnamefont {I.}~\bibnamefont
  {Almog}}, \bibinfo {author} {\bibfnamefont {Y.}~\bibnamefont {Sagi}},
  \bibinfo {author} {\bibfnamefont {G.}~\bibnamefont {Gordon}}, \bibinfo
  {author} {\bibfnamefont {G.}~\bibnamefont {Bensky}}, \bibinfo {author}
  {\bibfnamefont {G.}~\bibnamefont {Kurizki}},\ and\ \bibinfo {author}
  {\bibfnamefont {N.}~\bibnamefont {Davidson}},\ }\bibfield  {title} {\bibinfo
  {title} {Direct measurement of the system–environment coupling as a tool
  for understanding decoherence and dynamical decoupling},\ }\href
  {https://doi.org/10.1088/0953-4075/44/15/154006} {\bibfield  {journal}
  {\bibinfo  {journal} {Journal of Physics B: Atomic, Molecular and Optical
  Physics}\ }\textbf {\bibinfo {volume} {44}},\ \bibinfo {pages} {154006}
  (\bibinfo {year} {2011})}\BibitemShut {NoStop}%
\bibitem [{\citenamefont {Felinto}\ \emph {et~al.}(2005)\citenamefont
  {Felinto}, \citenamefont {Chou}, \citenamefont {de~Riedmatten}, \citenamefont
  {Polyakov},\ and\ \citenamefont {Kimble}}]{Felinto}%
  \BibitemOpen
  \bibfield  {author} {\bibinfo {author} {\bibfnamefont {D.}~\bibnamefont
  {Felinto}}, \bibinfo {author} {\bibfnamefont {C.~W.}\ \bibnamefont {Chou}},
  \bibinfo {author} {\bibfnamefont {H.}~\bibnamefont {de~Riedmatten}}, \bibinfo
  {author} {\bibfnamefont {S.~V.}\ \bibnamefont {Polyakov}},\ and\ \bibinfo
  {author} {\bibfnamefont {H.~J.}\ \bibnamefont {Kimble}},\ }\bibfield  {title}
  {\bibinfo {title} {Control of decoherence in the generation of photon pairs
  from atomic ensembles},\ }\href {https://doi.org/10.1103/PhysRevA.72.053809}
  {\bibfield  {journal} {\bibinfo  {journal} {Phys. Rev. A}\ }\textbf {\bibinfo
  {volume} {72}},\ \bibinfo {pages} {053809} (\bibinfo {year}
  {2005})}\BibitemShut {NoStop}%
\bibitem [{\citenamefont {Suarez}\ \emph {et~al.}(2022)\citenamefont {Suarez},
  \citenamefont {Wolf}, \citenamefont {Weiss},\ and\ \citenamefont
  {Slama}}]{Suarez}%
  \BibitemOpen
  \bibfield  {author} {\bibinfo {author} {\bibfnamefont {E.}~\bibnamefont
  {Suarez}}, \bibinfo {author} {\bibfnamefont {P.}~\bibnamefont {Wolf}},
  \bibinfo {author} {\bibfnamefont {P.}~\bibnamefont {Weiss}},\ and\ \bibinfo
  {author} {\bibfnamefont {S.}~\bibnamefont {Slama}},\ }\bibfield  {title}
  {\bibinfo {title} {Superradiance decoherence caused by long-range
  rydberg-atom pair interactions},\ }\href
  {https://doi.org/10.1103/PhysRevA.105.L041302} {\bibfield  {journal}
  {\bibinfo  {journal} {Phys. Rev. A}\ }\textbf {\bibinfo {volume} {105}},\
  \bibinfo {pages} {L041302} (\bibinfo {year} {2022})}\BibitemShut {NoStop}%
\end{thebibliography}
%
\end{document}